\documentclass[12pt]{article}
\usepackage{amsfonts,amsmath,amsthm,amssymb,verbatim,bm,bbm,makecell}
\usepackage{graphicx}
\usepackage{setspace,fullpage,color}
\usepackage[longnamesfirst]{natbib}
\usepackage[colorlinks,pagebackref=false]{hyperref}\usepackage[pdftex,letterpaper,left=.95in,top=1in,bottom=1in,right=.95in]{geometry}
\usepackage{titlesec}
\usepackage{enumitem}
\setlist{noitemsep,parsep=6pt,partopsep=0pt,topsep=0pt}
\usepackage{xparse}
\usepackage{mathtools}
\usepackage{centernot}
\usepackage{multibib} 
\newcites{supp}{Supplementary References}

\titlespacing\section{0pt}{10pt plus 2pt minus 2pt}{4pt plus 2pt minus 2pt} %
\titlespacing\subsection{0pt}{6pt plus 2pt minus 2pt}{2pt plus 2pt minus 2pt} %
\titlespacing\subsubsection{0pt}{6pt plus 2pt minus 2pt}{0pt plus 2pt minus 2pt} %
\titlespacing{\paragraph}{%
  0pt}{%
  0.5\baselineskip}{%
  1em}%

\providecommand{\U}[1]{\protect\rule{.1in}{.1in}}
\definecolor{dark-red}{rgb}{0.4,0.15,0.15}
\definecolor{dark-blue}{rgb}{0.15,0.15,0.75}
\definecolor{medium-blue}{rgb}{0,0,0.5}

\let\oldfootnote\footnote
\renewcommand\footnote[1]{\oldfootnote{\hspace{.4mm}#1}}
\makeatletter
  \renewcommand\@seccntformat[1]{\csname the#1\endcsname.{\hskip.7em\relax}} 
\makeatother
\newtheorem{theorem}{Theorem}

\newtheorem{condition}{Condition}
\newtheorem{corollary}{Corollary}
\newtheorem{lemma}{Lemma}
\newtheorem{proposition}{Proposition}

\theoremstyle{remark}
\newtheorem{remark}{Remark}
\newtheorem{examplex}{Example}
\newenvironment{example}
  {\begin{examplex}}
  {\hfill\qed
  \end{examplex}}

\theoremstyle{definition}
\newtheorem{definition}{Definition}

\renewenvironment{proof}[1][Proof]{\noindent\textbf{#1.} }{\hfill \ \textsquare}

\hypersetup{
pdftitle={Information Revelation in Constant-Sum Games},
pdfauthor={Kartik, Squintani, Tinn},
citecolor=dark-blue,
bookmarksnumbered=true,
    urlcolor=dark-red,
    linkcolor=dark-red
}
\def\E{\mathbb{E}}
\def\Reals{\mathbb{R}}
\def\Normal{\mathcal{N}}

\def\0{\mathbf{0}}
\newcommand{\ind}{\mathbbm{1}}

\def\citepos#1{\citeauthor{#1}'s (\citeyear{#1})}
\NewDocumentCommand{\citeposref}{O{} m}{\citeauthor{#2}'s (\citeyear{#2}, #1)}
\renewcommand{\epsilon}{\varepsilon}
\DeclareMathOperator{\sign}{sign}
\DeclareMathOperator{\Var}{Var}
\DeclareMathOperator{\Cov}{Cov}

\DeclareMathOperator{\Be}{Beta}
\DeclareMathOperator{\co}{co}

\newcommand{\norm}[1]{\left\lVert#1\right\rVert}
\DeclarePairedDelimiter{\abs}{\lvert}{\rvert}

\newcommand{\market}{m}

\newcommand{\appendixref}[1]{\hyperref[#1]{Appendix \ref{#1}}}

\begin{document}

\title{\Large{\textcolor{dark-blue}{%
Information Revelation in Constant-Sum Games:\\[5pt] Elections and Beyond}}\thanks{%
Earlier versions of this paper were circulated under the title ``Information Revelation and Pandering in Elections''. For helpful comments and discussion, we thank Nageeb Ali, Scott Ashworth, Cole Wittbrodt, 
Kfir Eliaz, Andrea Galeotti, Christian Hellwig, Matias Iaryczower,
Alessandro Lizzeri, Tianhao Liu, Massimo Morelli, David Rahman, Andres Santos, Jesse Shapiro, Bal\'{a}zs
Szentes, Olivier Tercieux, anonymous referees, the Co-Editor (Bruno Strulovici), as well as several seminar and conference audiences.
We are especially grateful to Afonso Gon\c{c}alves da Silva and Johannes H\"{o}%
rner. Kartik acknowledges, with gratitude, financial support from the Sloan
Foundation, the National Science Foundation (Grant SES-115593), and the
hospitality of and funding from the University of Chicago Booth School of
Business during a portion of this research.}}
\author{Navin Kartik\thanks{%
Yale University, Department of Economics. Email: \href{mailto:nkartik@gmail.com}%
{\texttt{nkartik@gmail.com}}. The author was affiliated with Columbia University during most of the work on this paper.} \and Francesco Squintani\thanks{%
University of Warwick, Department of Economics. Email: \href{mailto:F.Squintani@warwick.ac.uk}%
{\texttt{F.Squintani@warwick.ac.uk}}.} \and Katrin Tinn\thanks{%
McGill University, Desautels Faculty of Management. Email: \href{katrin.tinn@mcgill.ca}%
{\texttt{katrin.tinn@mcgill.ca}}}}
\date{\today \\ \small{(First version: March 2012)}}
\maketitle

\begin{abstract}
\noindent %

Do elections aggregate the private information of office-motivated candidates? Our answer stems from a general result for two-player constant-sum Bayesian games with type-independent payoffs. %
Under a ``completeness'' statistical condition, every ``identifiable'' equilibrium is an ex-post equilibrium.
Applied to Downsian elections, the ex-post property implies a sharp bound on information aggregation: equilibrium voter welfare is at best equal to the efficient use of a single candidate's information. In canonical specifications, politicians may ``anti-pander'' (overreact to their information), whereas some degree of pandering would be socially beneficial. We discuss other applications of the ex-post result.
\end{abstract}

\allowdisplaybreaks

\thispagestyle{empty}

\newpage

\setcounter{page}{1} \parskip=0.1in \setlength{\baselineskip}{20pt}

\setcounter{page}{1} \parskip=0.1in \setlength{\baselineskip}{21pt}

\section{Introduction\label{sec:intro}}

For representative democracy to be effective, voters must select representatives whose policies enhance their welfare. A challenge is that citizens are often poorly informed on policy issues, as posited by \citet{Downs:57} in his ``rational ignorance''\ hypothesis and since supported by numerous studies starting with \citet{CCMS:60}. Political candidates, by contrast, devote substantial resources and have broad access to policy experts and think tanks.  Politicians can convey their information to the electorate through their electoral campaigns, and in particular, through their policy positions. Indeed, there is evidence that voters learn and/or refine their views during elections.\footnote{\citet{LPP:23} provide large-scale cross-country evidence that a substantial share of voters decide late in campaigns, consistent with meaningful voter learning. Earlier work includes experiments on deliberative polling \citep{Fishkin:97}, studies on the effects of information on voters' opinions \citep{Zaller:92,Althaus:98,Gilens:01}, work on framing in polls \citep{SP:81}, and experiments on priming \citep{IK:87}.} But when office-seeking politicians choose their positions strategically, how effectively do elections aggregate their information?

One prevalent view is that elections function well because even office-seeking politicians are impelled to choose policies that promote voters' interests. Indeed,  \citet[p.~1400]{Wittman:89} influentially argued that political competition benefits the electorate because ``there are returns to an informed political entrepreneur from providing the information to the voters, winning office, and gaining the [...] rewards of holding office.'' Concurrently, however, there are also concerns---raised both in popular circles and in academic work that we discuss subsequently---that competitive pressures drive politicians to pander to voters' opinions rather than provide valuable information.  After all, the argument goes, it is hard to win an election by campaigning on policies with recondite merits; a politician is better off simply promising to do whatever voters believe is best from the outset. Pandering is viewed as inefficient because it would lead to policies that are excessively distorted toward the voters' less-informed opinions.

Our paper (re-)assesses the efficiency of elections when office-seeking politicians possess private information about policy consequences. \autoref{sec:main} lays out an extension of the canonical Downsian model of elections \citep{Downs:57,Hotelling:29}. Our framework is quite general, but we maintain the Downsian assumption of two candidates making policy commitments to maximize their probability of winning the election. The key twist is that each politician has (imperfect) private information about policy consequences. In other words, they each have information about which policy would be best for a representative or median voter---hereafter, ``the voter.''

We show that Downsian elections are fundamentally limited in their ability to aggregate candidates' private information.  Under reasonable conditions, \autoref{thm:welfare} deduces that in any equilibrium, holding fixed the politicians' equilibrium strategies, voter welfare---ex-ante expected utility---equals the welfare from (hypothetically) electing the same candidate regardless of platforms. That implies a tight upper bound: equilibrium voter welfare is no higher than what can be obtained based on one politician's information alone 
(\autoref{thm:competition}).

The key to establishing those conclusions is a general property of a class of constant-sum Bayesian games, studied in \autoref{sec:zerosum}. Specifically, consider any two-player constant-sum Bayesian game with type-independent payoffs.\footnote{In the electoral context, the two players are the politicians; office motivation (each candidate is maximizing their probability of winning) means that no matter the voter's strategy, the politicians are engaged in a constant-sum game. Their private information, while certainly relevant to the voter, does not directly affect the politicians' payoffs.} Our main theoretical result, \autoref{thm:expost-general},  says that under a general completeness condition on the distribution of types, any equilibrium in which at least one player uses an ``identifiable'' strategy---for example, a pure strategy---must be an ex-post equilibrium. That is, even after observing the opponent’s action, a player must be indifferent among all his own on-path actions. This ex-post property substantially limits the scope for how much private information a player's actions can reveal in purely competitive settings, such as a Downsian election.

To understand politicians' strategic incentives in more detail, we turn in \autoref{sec:pandering} to a one-dimensional normal-quadratic specification of the Downsian election. We assume there that the best policy for the voter---the ``state'' of the world---is drawn from a normal distribution; each candidate's private signal is the true state plus noise that is also normally distributed; and the voter's payoff is a quadratic loss function of the distance between the chosen policy and the state.

For that specification, we explain why it is not an equilibrium for each politician to propose a policy that is best for the voter based on his own information, i.e., to use an ``unbiased strategy'' (in which case the election would aggregate more than one politician's information). We show that, perhaps contrary to intuition, politicians would have an incentive to deviate by ``anti-pandering''---\emph{overreacting} to their private information---as the rational voter would elect the more extreme politician under unbiased strategies. The voter would do so because each politician's estimate of the state based on his own signal places more weight on the prior than the voter's estimate after learning both politicians' signals.\footnote{\label{fn:statistical}\citet{GS:09} and \citet{RouxSobel:12} also identify this implication of Bayesian updating in a non-strategic group decision-making context.}

Building on the above logic, we identify in \autoref{thm:FRE} a symmetric equilibrium that features anti-pandering by both politicians.  In this equilibrium, politicians choose different platforms with probability one, yet---regardless of their platforms---are elected with equal probability. Although there are other equilibria, the anti-pandering equilibrium shows starkly that office motivation need not induce pandering (or underreaction to private information). Further, in terms of welfare, we show in \autoref{p:benevolent-best-weaker} that a suitable degree of disequilibrium pandering would actually benefit the voter, contrary to perceptions that pandering is always harmful. 

Although our main economic application concerns Downsian elections---or any equivalent setting in which two agents compete for their proposals to be selected by a decision-maker---the abstract ex-postness result of \autoref{thm:expost-general} has broader relevance. In \autoref{sec:dualcompetition} we develop another application, in which two firms compete for both private market share and a government action (e.g., procurement). The firms have private information about a fundamental that matters for the government's optimal allocation, for example the relative social value of their products. Although the government might benefit from learning about the fundamental from the firms' choices, we show that the ex-post property implied by our key statistical condition often precludes such benefit. We also discuss how the framework in \autoref{sec:dualcompetition} has broader applicability.

\paragraph{Related Literature.}

Our \autoref{thm:expost-general} and its ingredient \autoref{lem:zerosum_general} relate to work on the equilibrium properties of two-player constant-sum games (hereafter just ``constant-sum games''), which dates back to \citet{VN:28}. Our results go beyond equilibrium payoff uniqueness or interchangeability \citep{Nash:51}, by establishing, under some conditions, an ex-post property for a class of constant-sum games: Bayesian games with type-independent payoffs. Ex-postness is not a general equilibrium property of constant-sum games; simply consider ``matching pennies''.  While the class of games \autoref{lem:zerosum_general} or \autoref{thm:expost-general} apply to is (very) restricted, we demonstrate relevant economic applications. There are two papers we are aware of with closely-related results.\footnote{As part of their study of robust implementation, \citet[Theorem 4, part 1]{PS24} observe that if two agents' payoffs are state-independent (but not necessarily constant sum), then any equilibrium of the game with no information about the state can also be supported as an equilibrium when agents observe the state. The reason is that the state is merely a correlating device. However, this observation only addresses some of the equilibria when agents observe the state (which is sufficient for their purposes), and moreover, these equilibria need not be ex-post equilibria. Our \autoref{thm:expost-general}'s conclusion of ex-postness of all (identifiable) equilibria owes to its constant-sum and statistical-completeness assumptions.}   First, \citet[Proposition 3.8]{Viossat:06} derives certain properties of correlated equilibria of complete-information constant-sum games, which, as we explain after \autoref{lem:zerosum_general}, is connected to our lemma. He does not have an analog to \autoref{thm:expost-general}. 

Second, \citet{Kattwinkel22} study mechanisms without transfers when two agents, each with a finite set of types, have type-independent and opposing preferences over a binary allocation. Their Proposition 1 characterizes incentive compatibility of direct mechanisms, and their Proposition 3 (part 2) shows that under a full-rank condition, only constant mechanisms are incentive compatible. For finite type sets, these results are related to our \autoref{lem:zerosum_general} and \autoref{thm:expost-general}, as elaborated in \autoref{sec:zerosum}. We study more general games rather than just direct mechanisms; moreover, a treatment of infinite type sets is valuable for applications, including our main electoral one, whose leading specification (\autoref{sec:pandering}) has normally distributed types.

With regard to our main application, there is a small prior literature on electoral competition when candidates have policy relevant private information.\footnote{There are also models in which candidates have private information that is not policy relevant for voters, e.g., about the location of the median voter \citep{OS:06,BDS:07,BDS:09}.} \citet{HL:03} illustrate why candidates may have an incentive to pander to the electorate's prior belief; their setting is one with binary policies, binary states, and binary signals. We find that in our richer setting, the opposite may be true for a broad class of information structures. Plainly, with binary policies, one cannot see the logic of why and how candidates may wish to overreact to private information. \citet{Loertscher:10} maintains the binary signal and state structure, but introduces a continuum policy space. His results are more nuanced, but at least when signals are sufficiently precise, the conclusions are similar to those of \citet{HL:03}.\footnote{In the \hyperref[app:beta-bernoulli]{Supplementary Appendix}, we show how overreaction or anti-pandering arises in a binary-signal model specification when the policies and the state lie in the unit interval. That specification permits a closer comparison with \citet{HL:03} and \citet{Loertscher:10}.}

\citet{LVdS:04} show that if voters in the \citet{HL:03} model are endowed with sufficiently precise private information about the policy-relevant state, then there are equilibria in which candidates fully reveal their private information; see also \citet{Klumpp:11} and \citet{Gratton:14}. By contrast, we are interested in settings in which  there is little information voters have that candidates do not. 

The anti-pandering equilibrium of our normal-quadratic model specification
provides a new perspective on the classic issue of policy divergence. Unlike
some other prevalent explanations (e.g., ideologically-motivated candidates
with uncertainty about voter preferences, as in \citet{Wittman:83} and %
\citet{Calvert:85}), anti-pandering features office-motivated politicians
diverging in order to maximize support from a risk-averse voter whose
ideology is known.\footnote{%
Other explanations for divergence include those based on increasing turnout %
\citep{Glaeser:05}, campaign contributions \citep{%
Campante:11},
valence asymmetries \citep{Groseclose:01,AP:02}, signaling character,
competence, or related mechanisms \citep{CW:07,KM:07,Honryo:18}, or more
than two candidates \citep{Palfrey:84}.}

Building on earlier versions of the current paper, \citet{MOS:20} introduce
confirmation bias for voters in a continuum-policy ternary-state model. They
find that confirmation bias can reduce equilibrium anti-pandering.

\citet{Schultz:96} studies a model in which two candidates are perfectly informed about the policy-relevant state but are policy motivated. He finds that when the candidates' ideological preferences are sufficiently extreme, platforms cannot reveal the true state; however, because of the perfect information assumption, full revelation can be sustained when ideological preferences are not too extreme. \citet{Martinelli:01} and \citet{MM:02} derive further results with ideologically motivated candidates who are perfectly informed about a policy-relevant variable.

\citet{ABK:16} study a model related to our normal-quadratic specification, but with candidates who are policy motivated. We explain in \autoref{sec:elections-discussion} that our welfare result continues to hold, approximately, when the extent of policy motivation is small. \citet{ABK:16} show that when policy motivation looms large and candidates' ideologies are sufficiently similar to the voter's, equilibria can aggregate more information. Our papers are complementary.

There are various other settings in economics and political science in which distortions arise because agents wish to influence their principals' beliefs. In particular, electoral models often feature a single politician seeking to build a reputation for either competence \citep[e.g.,][]{CWHS:01} or aligned preferences \citep[e.g.,][]{MT:04}. While most such papers highlight the possibility of pandering---or even ``over-pandering''\ as in \citet{AES:13} and \citet{KvW:17}---anti-pandering arises in \citet{PS:96}, \citet{Levy:04}, and \citet{Bils23}.

Finally, we note that our applications illustrate that ex-post equilibrium or incentive constraints leave limited scope for information revelation among two purely competitive players. Although in a very different setting, that is reminiscent of negative results like \citet{JMMZ06}, who show that generically only constant mechanisms are ex-post implementable under interdependent values and multidimensional signals.

\section{Two-Player Constant-Sum Bayesian Games
\label{app:generalBayesiangames}}
\label{sec:zerosum}

Our applications are underpinned by a general result on two-player constant-sum Bayesian games with type-independent payoffs. This section develops that result.

\paragraph{Setting.}

There are two players, $A$ and $B$. Each player $i\in \{A,B\}$ has a private type $%
s_{i}\in S_{i}$, where $S_{i}$ is a nonempty standard Borel space (e.g., a Borel subset of a Euclidean space).\footnote{Throughout, any standard Borel space is equipped with a compatible Polish topology; when the space is Euclidean, this is the usual topology.} %
The type profile $(s_{A},s_{B})$ is drawn from a common-prior probability measure $F$ on $S_{A}\times
S_{B} $, whose marginals $F_{A}$ and $F_{B}$ have supports $S_{A}$ and $%
S_{B}$, respectively. We assume $F$ is absolutely continuous with respect to the product measure $F_{A}\otimes F_{B}$, i.e., any set of signal profiles that has probability zero when types are drawn independently from the marginals also has probability zero under $F$. This is automatic when $S_{A}$ and $S_{B}$ are finite; with continuous signals, it rules out cases such as perfect correlation. Writing $-i$ for the player different from $i$ as usual, we will denote by $F(\cdot \mid s_i)$ the regular conditional
distribution of $s_{-i}$ given $s_i$.\footnote{This exists and is unique almost everywhere (a.e., hereafter) because $S_A \times S_B$ is a standard Borel space \citep[pp.~229--230]{Durrett95}.}

After learning their types, players choose actions simultaneously; player $i$'s action is denoted $x_{i}\in
X_{i}$, where each $X_{i}$ is a nonempty standard Borel space.\footnote{As usual, each $x_i$ can also be interpreted as player $i$'s action plan in a sequential-move game.}
Player $i$'s (von-Neumann--Morgenstern) payoff is $u_{i}(x_{i},x_{-i})$, with $%
u_{A}\left( \cdot \right) +u_{B}\left( \cdot \right) =0$. So the game is constant sum, with the sum normalized to $0$, and types do not directly affect payoffs. (Nevertheless, the players' types may affect the payoffs of third parties, as in our subsequent applications.)
Assume payoffs are measurable and uniformly bounded: $\abs{u_{i}(\cdot )}\leq {M}$, for
some constant ${M}.$ We denote each player $i$'s space of mixed actions---randomizations over actions---by $\Delta (X_{i})$, with generic element $\xi _{i}$, and extend payoffs to $\left( \Delta (X_{A}),\Delta
(X_{B})\right) $ by linearity as usual, writing $u_{i}(\xi
_{i},\xi _{-i})$. A mixed strategy for player $i$ is a measurable map $\sigma
_{i}:S_{i}\rightarrow \Delta (X_{i})$. We study Bayes-Nash equilibria: $(\sigma^*_A,\sigma^*_B)$ is an equilibrium if for each player $i$, the mixed action $\sigma^*_i(\cdot\mid s_i)$ is optimal against $\sigma^*_{-i}$ for $F_i$-a.e.~$s_i$.

\paragraph{Identifiability conditions.}
\autoref{thm:expost-general} below requires the following statistical condition on the distribution of signals. %
As is common, we use notation like $F(g(s_{i})=0)$ as shorthand for $F\left(\{s_i:g(s_{i})=0\}\right)$.

\begin{condition}
\label{completeness_general} For any $i\in \{A,B\}$ and any bounded measurable function $%
g:S_{-i}\rightarrow \mathbb{R}$, it holds that
\begin{align}
\mathbb{E}_{s_{-i}}[g(s_{-i})\mid s_{i}]=0\text{\textit{ for $F_i$-a.e.~}}%
s_{i}\in {S}_{i} \implies F\left(g(s_{-i})=0\mid s_{i}\right) =1 \text{\textit{ for $F_i$-a.e.~}}%
s_{i}\in S_{i}.  \label{e:complete}
\end{align}
\end{condition}

\autoref{completeness_general} is the property of \emph{bounded completeness} \citep[e.g.,][p.~144]{Lehmann:86} of the family of conditional distributions $\{F(\cdot \mid s_i)\}_{s_i \in S_i}$, understood up to $F_i$-null sets: the only bounded function of $s_{-i}$ whose conditional expectation vanishes at almost every $s_i$ is the zero function almost surely (a.s., hereafter). Put differently, there is enough variation in the family of conditional distributions that distinct bounded functions of the opponent's type induce distinct conditional expectations.

To illustrate, consider $S_A=S_B=\{1,2\}$. If signals are perfectly correlated ($\Pr(s_A=s_B)=1$), then $\E[g(s_{-i})\mid s_i]=g(s_i)$, which is $0$ for both $s_i$ only if $g\equiv 0$; hence \autoref{completeness_general} holds. If instead signals are independent, any nonzero $g$ with mean zero has $\E[g(s_{-i})\mid s_i]=0$ for both $s_i$, and \autoref{completeness_general} fails.

More generally, when $S_A$ and $S_B$ are both finite, bounded completeness for player $i$ alone is equivalent to full column rank of the conditional-probability matrix $[F(s_{-i}^k \mid s_i^j)]_{j,k}$; requiring the condition for both players is equivalent to the corresponding joint-probability matrix having full row and column rank (which requires $|S_A|=|S_B|$). \citet{CM85,CM88} use this linear-independence notion of richness for full surplus extraction in mechanism design with finite type spaces. We provide a linear-independence characterization for infinite type spaces in \autoref{app:SLI}: ``strong'' linear independence is equivalent to the statistical notion of \emph{completeness}, which in infinite spaces slightly strengthens bounded completeness by admitting unbounded test functions.

Plainly, if either $\abs{S_A}>1$ or $\abs{S_B}>1$, then \autoref{completeness_general} is violated if the
types are independent. But we are interested in settings in which each player's type is informative about the other's, stemming from both being informative about some underlying ``state of the world''. In those contexts, we view \autoref{completeness_general} as a reasonable requirement.

\autoref{completeness_general} holds, in particular, in the canonical case of exponential-family signals. That is, for each $i\in\{A,B\}$: $S_i \subset \Reals^n$; the interior of $S_i$ has $F_i$-probability one; and, for $s_i$ in this interior, the conditional distribution of $s_{-i}\mid s_i$ is a regular exponential family with natural parameter $s_i$ and density of the form
\begin{equation}\label{e:exponential}
f(s_{-i}\mid s_i)
= \exp\!\big(s_i\cdot T(s_{-i})-\psi(s_i)\big)\,h(s_{-i}),
\end{equation}
where the sufficient statistic $T: S_{-i} \to \Reals^n$ is injective, the cumulant function $\psi: S_i \to \Reals$ is continuous, and $h: S_{-i} \to \Reals_{\geq 0}$ is the carrier density.\footnote{\label{fn:exponential-complete}To see why \autoref{completeness_general} holds, fix $i$ and a bounded measurable $g: S_{-i}\to\Reals$ with $\E[g(s_{-i})\mid s_i]=0$ for $F_i$-a.e.~$s_i$. Continuity of $\psi$ implies---via pointwise convergence of the densities and Scheff\'e's theorem---that the map $s_i \mapsto \E[g(s_{-i})\mid s_i]$ is continuous on the interior of $S_i$. Since $F_i$ has support $S_i$ and assigns probability one to the interior, the set of signals at which the expectation vanishes is dense in the interior; by continuity, the expectation vanishes on the entire interior. Now, when the natural parameter ranges over a set with nonempty interior, the family of distributions of the statistic $T(s_{-i})$ is complete \citep[Theorem~1, p.~142]{Lehmann:86}. Because $T$ is injective and the spaces are standard Borel, $g$ can be written as a bounded measurable function of $T(s_{-i})$, so this completeness yields $g(s_{-i})=0$ $F(\cdot \mid s_i)$-a.s.~for every interior $s_i$, and hence for $F_i$-a.e.~$s_i$.} This class includes a variety of widely-used continuous distributions with bounded and unbounded supports, such as normal, exponential, gamma, beta, chi-squared, and Dirichlet.  For finitely-supported distributions such as the binomial, \autoref{completeness_general} can instead be verified using its rank characterization.

\autoref{thm:expost-general} also requires the following analog of \autoref{completeness_general} on players' strategies. 

\begin{definition}
\label{identifiability}
Strategy $\sigma_i$ is \emph{identifiable} if for any bounded
and measurable $g:X_i\rightarrow \mathbb{R}$, it holds that
\begin{equation*}
\mathbb{E}_{x_i}\left[ g\left(x_{i}\right) \mid s_{i}\right] =0\text{ for
$F_i$-a.e.~} s_{i}\in S_{i}\implies \Pr \left(g\left(
x_{i}\right) =0 \mid s_{i}\right) =1\text{ for $F_i$-a.e.~}s_{i}\in S_{i},
\end{equation*}%
where the left-hand-side expectation and right-hand-side
probability are computed using $\sigma _{i}$.

\end{definition}

Any pure strategy $\sigma_i:S_i \to X_i$ is identifiable because in that case $\E_{x_i}[g(x_{i})\mid s_{i}]=g(\sigma_{i}(s_{i}))$. Mixed strategies can be identifiable as well; for instance, with two actions, any strategy that assigns an action different probabilities on two positive-probability sets of signals is identifiable. An example of a non-identifiable strategy is any non-pure strategy that does not vary with the player's signal; less obviously, a strategy can vary with the signal and still fail identifiability if its variation is too low-dimensional relative to the action space.\footnote{Take $X_i = \{a,b,c\}$ and $\sigma_i(\cdot \mid s_i) = (s_i/2, s_i/2, 1-s_i)$ with $s_i \in [0,1]$. Then $g(a) = 1$, $g(b) = -1$, $g(c) = 0$ satisfies $\E[g(x_i) \mid s_i] = 0$ for every $s_i$, even though $g \not\equiv 0$.}

We do not restrict attention to pure strategies because the force of \autoref{lem:zerosum_general} and   \autoref{thm:expost-general}, and hence their applications, turns on the set of equilibria they cover: confined to pure strategies, they would leave open whether players might reveal more information by randomizing. Some qualification is needed for \autoref{thm:expost-general} (see \autoref{eg:identifiability}); strategy identifiability is a natural condition whose role is transparent in the theorem's proof.

\paragraph{The result.} Since the game has type-independent payoffs, standard arguments for constant-sum games imply that all equilibria yield the same payoff vector $(U^*_A,U^*_B)$. We say that an equilibrium $\sigma ^{\ast }:= (\sigma _{A}^{\ast},\sigma _{B}^{\ast })$ is an \emph{ex-post equilibrium} if $u_{i}(x_{i},x_{-i})=U_{i}^{\ast }$ for each $i\in \{A,B\}$ and for $(\sigma ^{\ast },F)$-a.e.~pair $(x_{A},x_{B})$. In other words, in an ex-post equilibrium, no type of either player would have an incentive to  take any other on-path action even after learning the action played by the opponent.\footnote{To be precise, this interpretation is valid when a player's payoff is constant across all action pairs $(x_i,x_{-i})$ in which each of $x_i$ and $x_{-i}$ is on the equilibrium path, even if the pair itself is not. That property is in fact established in the proof of \autoref{thm:expost-general}. Note that our verbal discussion ignores probability-zero caveats; to lighten the exposition, we frequently omit such caveats outside of formal statements. We also stress that our notion of ex-postness concerns only \emph{on-path} actions; it does not preclude a player strictly preferring some off-path action after learning the opponent's action.} We also say that an equilibrium $\sigma^*$ is an \emph{identifiable equilibrium} if either $\sigma^*_A$ or $\sigma^*_B$ is identifiable. In particular, any equilibrium in which at least one player is playing a pure strategy is identifiable.

\begin{theorem}
\label{thm:expost-general} If \autoref{completeness_general} holds, then any identifiable equilibrium $\sigma ^{\ast }$ is an ex-post equilibrium.
\end{theorem}

The theorem says that, subject to \autoref{completeness_general}, in any identifiable equilibrium the players are indifferent over all the action profiles that are played in equilibrium---even though different types of a player may be playing different (distributions over) actions and hold different beliefs about the opponent's type. Only one player's strategy need be identifiable because, as the proof shows, that pins down his opponent's payoff after almost every on-path action pair; the constant-sum property then pins down his own payoff as well.

The theorem's proof requires the following lemma, which is of independent interest as it does not rely on either \autoref{completeness_general} or identifiability of strategies. The lemma says that in any equilibrium, all types of a player obtain the same interim expected payoff (which is independent of the equilibrium), and any type would obtain that interim payoff regardless of which action it plays among all the actions taken by some type of that player.

\begin{lemma}
\label{lem:zerosum_general} Let $
(U_{A}^{\ast },U_{B}^{\ast })$ denote the payoffs in every equilibrium, and let $\sigma ^{\ast }$ be some equilibrium. Then for each $i\in \{A,B\}$, for $(\sigma
_{i}^{\ast },F_{i})$-a.e.~$x_{i}$ and $F_{i}$-a.e.~$s_{i}$,
it holds that 
\begin{equation*}
U_{i}^{\ast }=\mathbb{E}_{x_{-i}}[u_{i}(x_{i},x_{-i})\mid s_{i}],
\end{equation*}
where the expectation is taken with respect to the measure induced by $%
\sigma _{-i}^{\ast }$ and $F$.
\end{lemma}

\begin{proof}
As is standard, say that a mixed action $\xi_i \in \Delta(X_i)$ \emph{secures} player $i$ the payoff $U_i\in \mathbb{R}$ if $u_i(\xi_i,\xi_{-i}) \geq U_i$ for all $\xi_{-i}\in \Delta(X_{-i})$. Fix an equilibrium $\sigma^*$. Let $\xi^*_i$ be the mixed action defined as the ex-ante measure over $X_i$ induced by the strategy $\sigma^*_i$. Since the game is constant sum with type-independent payoffs, $\xi^*_i$ secures player $i$ the payoff $U^*_i$.\footnote{The statement follows from standard logic for constant-sum games, which we detail for completeness. If $\xi^*_i$ does not secure $U^*_i$, then there is some mixed action $\xi_{-i}$ such that $u_i(\xi^*_i,\xi_{-i})<U^*_i$, or equivalently by the constant-sum property, $u_{-i}(\xi_{-i},\xi^*_i)>U^*_{-i}$. But then playing the constant strategy $\sigma_{-i}(s_{-i})=\xi_{-i}$ would be a profitable deviation for player $-i$ against $\sigma^*_i$.} This implies that for each $i$, conditional on $F_i$-a.e.~types $s_i$, player $i$'s interim equilibrium payoff in fact equals $U_i^*$.
We now observe that for $F_i$-a.e.~types $s_i$, the conditional distribution of the
opponent's actions induced by $\sigma^*_{-i}$ and $F$ secures the opponent $U^*_{-i}$; for if not, there would be some action that yields $s_i$ a payoff strictly larger than $U_i^*$. Hence, for $(\sigma^*_i,F_i)$-a.e.~$x_i$ and $F_i$-a.e.~$s_i$, it follows that $\mathbb{E}_{x_{-i}}[u_{i}(x_{i},x_{-i})\mid s_{i}]=U^*_i$; the expectation cannot be larger by the preceding observation, and it then cannot be smaller either because $\xi^*_i$ secures $U^*_i$.
\end{proof}

One way to appreciate the content of \autoref{lem:zerosum_general} is via its connection to \emph{correlated} equilibrium of \emph{complete-information} games. Consider a {complete-information} two-player constant-sum game $G$ with action spaces $X_{A}$ and $X_{B}$ and payoff functions $u_{A}$ and $u_{B}$ as above. Any (objective) {correlated} equilibrium of this game is a Bayes-Nash equilibrium of our Bayesian game with a suitably-defined information structure; conversely, any Bayes-Nash equilibrium of our game is a correlated equilibrium of $G$. It follows from \autoref{lem:zerosum_general} that if $\rho \in \Delta (X_{A}\times X_{B})$ is a correlated equilibrium of $G$ with payoffs $(\pi _{A},\pi _{B})$, then for any $i\in \{A,B\}$ and $\rho$-a.e.~$x_{i}$ and $x_{i}^{\prime }$, it holds 
that $\mathbb{E}_{x_{-i}}\left[u_{i}(x_{i}^{\prime },x_{-i})\mid x_{i}\right] =\pi _{i},$ and hence $x_{i}^{\prime }$ is a best response to $\rho (\cdot \mid x_{i})$. For finite games, this fact has been noted by \citet[Proposition 3.8]{Viossat:06}.\footnote{\label{fn:Viossat}Moreover, because of the ``conversely'' point noted earlier in the paragraph, if we restricted to finite types and actions, \citepos{Viossat:06} result could in turn be used to prove \autoref{lem:zerosum_general}. Indeed, that indirect approach was used in earlier versions of our paper \citep[Appendix A]{KST15} and by \citet[Lemma 1 and Proposition 1]{Kattwinkel22}.} 

\autoref{lem:zerosum_general} also relates to 
\citet[Proposition 1]{Kattwinkel22}. Our result is stronger for two reasons. First, we allow for infinite type sets. Second, our result applies to arbitrary action spaces and equilibrium strategies, not just direct mechanisms and truthful equilibria. When types may mix over their actions, \autoref{lem:zerosum_general} establishes that each type is indifferent among all the actions in any type's equilibrium mixture---not merely indifferent among all types' distributions. This is crucial for the proof of \autoref{thm:expost-general}, which we now turn to.

\begin{proof}[Proof of \autoref{thm:expost-general}]
Let $\sigma^*$ be an equilibrium in which player $-i$'s strategy is identifiable. Below, all expectations are with respect to the
measure induced by $(\sigma^*,F)$. Take any $(\sigma^*_i,F_i)$-full-measure subset $\hat X_i \subseteq X_i$ delivered by \autoref{lem:zerosum_general} and fix any $x_i \in \hat X_i$.  For $F_i$-a.e.~$s_i$, we have
\begin{align*}
U^*_i
&= \mathbb{E}_{x_{-i}}[u_i(x_i,x_{-i}) \mid s_i]
\quad {\text{\small by \autoref{lem:zerosum_general}}} \\[5pt]
&= \mathbb{E}_{s_{-i}}\!\left[
\mathbb{E}_{x_{-i}}[u_i(x_i,x_{-i}) \mid s_i,s_{-i}]
\mid s_i\right]
\quad {\text{\small by the law of iterated expectation}} \\[5pt]
&= \mathbb{E}_{s_{-i}}\!\left[
\mathbb{E}_{x_{-i}}[u_i(x_i,x_{-i}) \mid s_{-i}]
\mid s_i\right]
\quad {\text{\small because $x_{-i}$ is independent of $s_i$, conditional on $s_{-i}$.}}
\end{align*}

Now, applying \autoref{completeness_general} (just for one player, $i$) with $g(s_{-i})=\mathbb{E}_{x_{-i}}[u_i(x_i,x_{-i}) \mid s_{-i}] - U^*_i$, we get for $F_i$-a.e.~$s_i$ that 
\[
\mathbb{E}_{x_{-i}}[u_i(x_i,x_{-i}) \mid s_{-i}] = U^*_i
\quad \text{ for } F(\cdot\mid s_i)\text{-a.e.~} s_{-i}.
\]
Integrating over $s_i$ with respect to $F_i$ and using the law of total probability yields the above equality for $F_{-i}$-a.e.~$s_{-i}$.

It then follows from the identifiability of $\sigma^*_{-i}$, applied with $g(x_{-i})=u_i(x_i,x_{-i})-U_i^*$, that for $F_{-i}$-a.e.~$s_{-i}$ we have 
\[u_i(x_i,x_{-i}) = U^*_i \quad \text{ for } \sigma^*_{-i}(\cdot \mid s_{-i})\text{-a.e.~} x_{-i}.\]
Integrating over $s_{-i}$ with respect to $F_{-i}$  and using the law of total probability yields the above equality for $\left(\sigma_{-i}^*, F_{-i}\right)$-a.e.~$\left(x_{-i}, s_{-i}\right)$.

Let $\mu_{A}$ and $\mu_B$ denote the ex-ante distributions of the players' actions. The set $\hat X_i$ has $\mu_i$-measure one. Since $x_i$ was arbitrary in $\hat X_i$, we have $u_i(x_i,x_{-i})=U^*_i$ for $\mu_A\otimes\mu_B$-a.e.~$(x_A,x_B)$. Since $F$ is absolutely continuous with respect to $F_A\otimes F_B$, the joint distribution of $(x_A,x_B)$ induced by $(\sigma^*,F)$ is absolutely continuous with respect to $\mu_A\otimes\mu_B$,\footnote{\label{fn:domination}For measurable $N\subseteq X_A\times X_B$ with $(\mu_A\otimes\mu_B)(N)=0$, let $$q_N(s_A,s_B):=\int\!\!\int \ind\{(x_A,x_B)\in N\}\,\sigma^*_A(\mathrm dx_A\mid s_A)\,\sigma^*_B(\mathrm dx_B\mid s_B).$$ Then $\int q_N\,\mathrm d(F_A\otimes F_B)=(\mu_A\otimes\mu_B)(N)=0$, so $q_N=0$ $F_A\otimes F_B$-a.e.\ and hence $F$-a.e.; $N$ therefore has probability zero under the joint distribution.} so $u_i(x_i,x_{-i})=U^*_i$ holds $(\sigma^*,F)$-a.e. The analogous equality for player $-i$ follows from the game being constant sum.
\end{proof}

The following two examples show that neither \autoref{completeness_general} nor the qualification of identifiability can be dropped from \autoref{thm:expost-general}. Both examples are based on a ``matching pennies'' payoff structure.

\begin{example}
\label{eg:completeness}
Consider $X_A=X_B=S_A=S_B=\{0,1\}$, $u_A(x_A,x_B)=\ind\{x_A=x_B\}$, and $F$ the uniform distribution. \autoref{completeness_general} fails because $s_{A}$ and $s_{B}$ are independent. There is an identifiable equilibrium in which each player $i$ plays the pure strategy $s_i \mapsto s_i$. This is, however, not an ex-post equilibrium.
\end{example}

\begin{example}
\label{eg:identifiability}Now consider a complete-information variant of the previous example. There are singleton type sets, $\abs{S_A}=\abs{S_B}=1$; trivially, \autoref{completeness_general} holds. Actions and payoffs are as in \autoref{eg:completeness}. 
The equilibrium in which both players uniformly randomize over their two actions is not an ex-post equilibrium; these mixed strategies are not identifiable, and hence the equilibrium is not identifiable.
\end{example}

Note that \autoref{lem:zerosum_general} applies to both examples. In particular, in the equilibrium of \autoref{eg:completeness}, neither type of a player is playing an action that secures the equilibrium payoff; nevertheless, each type has the same interim payoff and is indifferent between its equilibrium action and the equilibrium action of the other type.

We close this section by fleshing out one implication of \autoref{thm:expost-general} that is useful for applications. Let $\Omega$ be some set of outcomes and $w:X_A\times X_B\to \Omega$ an outcome function. Assume that preferences depend on only the outcome, i.e., there is some $\tilde u_i:\Omega \to \Reals$ such that  $u_i(x_A,x_B)=\tilde u_i(w(x_A,x_B))$ for all $(x_A,x_B)$. Say that there are \emph{strict preferences over outcomes} if each $\tilde u_i$ is injective, and say that an equilibrium $\sigma^*$ has a \emph{single outcome} $\omega^*$ if $w(x_A,x_B)=\omega^*$ for $(\sigma^*,F)$-a.e.~$(x_A,x_B)$.  If there are strict preferences over outcomes, then plainly an ex-post equilibrium must have a single outcome. Hence, the following result follows directly from \autoref{thm:expost-general}.

\begin{corollary}
\label{cor:singleoutcome}
Assume strict preferences over outcomes. If \autoref{completeness_general} holds, then any identifiable equilibrium has a single outcome.
\end{corollary}

In fact, since the corollary holds for an arbitrary outcome space and function, it is equivalent to \autoref{thm:expost-general}. For, if we define outcomes as the utility-equivalence classes of action profiles and the outcome function as mapping each action profile into its equivalence class, then we have strict preferences over outcomes by construction, and a single outcome corresponds exactly to ex-postness.

As detailed in \autoref{app:Kattwinkel},  \autoref{cor:singleoutcome} implies \citepos{Kattwinkel22} Proposition 3, part 2. In a direct-mechanism setting with binary  allocations that two agents with a finite number of types have opposed preferences over, those authors show that incentive compatibility requires a constant allocation probability, so long as the type distribution has full rank.

In the remainder of the paper, we use \autoref{thm:expost-general}/\autoref{cor:singleoutcome} to study information revelation and aggregation in some economic settings.

\section{Information Aggregation and Pandering in Elections}
\label{sec:main}

This section studies a model of a Downsian election with
informed candidates. 

\paragraph{Model.}

We consider an electorate that is represented in reduced-form by a single voter. The voter can be viewed as a median voter when one exists, as elaborated in the next paragraph. Alternatively, she could be the representative of a group or simply a single decision maker, for instance a manager choosing between competing proposals \citep[cf.~][]{ABK:16}. The voter's preferences depend upon the implemented policy $x\in X$ and an
unknown state of the world $\theta \in \Theta $, where both $X$ and $\Theta $
are standard Borel spaces. The dimensionality of $X$ and $\Theta$ plays no role in our arguments, and the generality improves the fit for some applications, particularly when the voter is a single decision maker. For instance, she may choose a bundle of policies, say spending and regulation, with the state describing the socially appropriate level of each.

The state is drawn from a probability measure $F_{\theta }$. The voter's preferences are represented by a von-Neumann--Morgenstern utility function $u:X\times \Theta \rightarrow \mathbb{R}$. We assume integrability: $\E[\abs{u(x, \theta)}]<\infty$ for all $x \in X$, where the expectation is under the prior $F_\theta$. We also assume there is a utility-maximizing policy in each state and an expected-utility-maximizing policy under the prior. A leading example that we will return to is the (one-dimensional) quadratic
loss function: $X,\Theta \subset \mathbb{R}$ and $u(x,\theta )=-(x-\theta)^{2}$, with $F_\theta$ having finite second moment. For this example, the median voter interpretation is valid for an electorate whose members $j$ have utilities $u_j(x,\theta)=-(x-\theta-b_j)^2$ for parameters $b_j\in \Reals$ (with median normalized to $0$). More generally, the median interpretation requires a suitable single-crossing property under uncertainty.\footnote{Specifically, it is sufficient if the electorate has ordered voter types $j\in J$ and utilities $\{u_j(x,\theta)\}_{j\in J}$ such that for each policy pair $x$ and $x'$, the utility difference $u_j(x, \theta)-u_j(x^{\prime}, \theta)$ is single-crossing in the type $j$ for each state $\theta$, and these functions satisfy signed-ratio monotonicity across states \citep{QS12}. In turn, that is assured if the utilities satisfy \citepos{KLR23} single-crossing expectational differences in $((x,\theta);j)$.}

There are two electoral candidates, $A$ and $B.$ Given the state $\theta $, each candidate $i\in \{A,B\}$ privately observes a signal $s_{i}\in S_{i}$, where $S_{i}$ is a closed subset of $\mathbb{R}^{n},$ with $n\geq1$. The joint conditional cumulative distribution of $(s_A,s_B)\in S_A \times S_B$ is denoted by $F_{s_A,s_B \mid \theta}$, and the conditional marginal for each candidate $i$ by $F_{s_{i} \mid \theta }$. The measure $F_{\theta }$ and distributions $F_{s_A,s_B \mid \theta }$ induce a joint cumulative distribution $F_{s_{A},s_{B}}$ of signal profiles unconditional on the state, with marginals $F_{s_{A}}$ and $F_{s_{B}}$. We assume that $F_{s_{A},s_{B}}$ is absolutely continuous with respect to the product of the marginals, $F_{s_{A}}\otimes F_{s_{B}}$, and that for each candidate $i$, the support of $F_{s_i}$ is $S_i$. We also assume that for each candidate $i$ and signal $s_i\in S_i$, there is a voter-optimal policy:   $\max_{x\in X} \E_\theta [u(x,\theta) \mid s_i]$ exists, and these maximizers admit a measurable selection with finite ex-ante expected utility.

Our results will require that the joint cumulative distribution $F_{s_{A},s_{B}}$ satisfy \autoref{completeness_general}. This is a reasonable requirement when both signals $s_{A}$ and $s_{B}$ are informative about the
state; in particular, following the discussion after \autoref{completeness_general}, typical exponential families of signal distributions satisfy the requirement. A leading example that we will return to is the (one-dimensional) \emph{normal-normal} structure: $\Theta=S_A=S_B =\Reals$, $\theta \sim \Normal(0,1/\alpha)$, i.e., the state is normally distributed with mean 0 and precision $\alpha\in \Reals_{>0}$, and conditional on the state $\theta$, each candidate $i$'s signal is drawn independently from the normal distribution $\Normal(\theta,1/\beta_i)$ with precision parameter $\beta_i \in \Reals_{>0}$.\footnote{In this case, it is routine to verify that conditional on signal \( s_i \), the distribution of signal \( s_{-i} \) is normal with mean
$
\frac{\beta_i}{\alpha + \beta_i} s_i
$
and variance $\sigma^2:=
\frac{1}{\alpha + \beta_i} + \frac{1}{\beta_{-i}}$.
Hence, \autoref{e:exponential} holds with
\[
T(s_{-i}) = \frac{1}{\sigma^2} \cdot \frac{\beta_i}{\alpha + \beta_i} s_{-i}, \quad
\psi(s_i) = \frac{1}{2\sigma^2} \left( \frac{\beta_i}{\alpha + \beta_i} s_i \right)^2, \quad
\text{and} \quad
h(s_{-i}) = \frac{1}{\sigma \sqrt{2\pi}} e^{- \frac{1}{2\sigma^2} s_{-i}^2}.
\]
}
The case of $\beta_A\neq \beta_B$ captures candidates having access to information of different quality. In general, $F_{s_A,s_B}$ can satisfy \autoref{completeness_general} even with signals being positively (or negatively) correlated conditional on the state.

After privately observing their signals, the candidates simultaneously
choose their platforms $x_{A}$ and $x_{B}$ from the policy space $X$, with
the objective of maximizing their respective probabilities of winning the
election.\footnote{%
We assume that both candidates can choose from the same set of platforms for
notational simplicity. Our analysis in this section would hold equally well
if each candidate $i$ can only choose platforms from some subset $X_i\subset
X$. One could use $X_A\neq X_B$ to capture asymmetries between the
candidates, e.g., if there is an incumbent and a challenger, and the
incumbent's history precludes him from choosing certain policies.} Upon
observing the platforms $(x_A,x_B),$ the voter updates her belief about the
state $\theta$ and then elects the candidate whose platform provides the highest
expected utility. The elected candidate $i\in \{A,B\}$ implements his
platform $x_i$. Platforms are thus policy commitments in the Downsian
tradition. Candidates are expected utility maximizers, with the elected candidate obtaining a utility of $1$ and the other candidate $0$. Hence, candidates are purely office motivated. All aspects of the model except the candidates' private signals
are common knowledge.

\paragraph{Strategies, Equilibria, and Welfare.}

A pure strategy for a candidate $i$ is a measurable function $y_{i}:S_{i}\rightarrow X$, with $y_{i}(s_{i})$ the platform chosen by $i$ when his signal is $s_{i}$. A strategy for the voter is a measurable function $w_{A}:X^{2}\rightarrow[0,1]$, where $w_{A}(x_{A},x_{B})$ represents the probability with which candidate $A$ is elected when the platforms are $x_{A}$ and $x_{B}$. Candidate $B$ is elected with the complementary probability $w_B(x_B,x_A):=1-w_A(x_A,x_B)$. 

We study (weak) perfect Bayesian equilibria $(y_{A},y_{B},w_{A})$ of the electoral game in which candidates play pure strategies---hereafter, simply  \emph{equilibria}.\footnote{Our leading specifications---such as the normal-normal structure---have continuous signals with atomless distributions, in which case it is salient to focus on equilibria with pure candidate strategies. \autoref{p:UD} assures that such equilibria exist regardless of the model specification. Notwithstanding, we discuss equilibria in which candidates may mix in \autoref{sec:elections-discussion}.} Hence, as noted in \autoref{app:generalBayesiangames}, the candidates' strategies are identifiable in these equilibria. The voter elects candidate $i$ if $x_{i}$ is strictly preferred to $x_{-i}$. We allow the voter to randomize arbitrarily when indifferent. For the median voter interpretation, one may want to insist on uniform randomization;  our results would be unaffected by this requirement, modulo one caveat noted in \autoref{fn:50-50}.

The notion of welfare we adopt is the voter's ex-ante expected utility, which we denote $v(y_A,y_B,w_A)$ as a function of the strategies. We restrict attention to candidate strategies satisfying $\E[\abs{u(y_{i}(s_{i}),\theta )}]<\infty$, so that voter welfare is well defined.

\paragraph{Policy Commitment.}In the Downsian tradition, our model assumes that candidates make commitments to the policies they would implement if elected.\footnote{See \citet{OS:96}, \citet{BC:97} and subsequent work for non-Downsian ``citizen-candidate'' models.} In reality, while commitment may be imperfect, some degree of it is plausible  and valuable to candidates; in their meta-study of earlier research, \citet{PC:09} conclude that around 67\% of campaign promises have historically been kept. The theoretical literature has proposed multiple rationales for commitment, most prominently that of re-election concerns \citep{Alesina:88}. Alternatively, if there is uncertainty about a candidate's quality of information and candidates have reputation concerns (perhaps because of re-election motives), then ``flip flopping'' or ``vacillating'' may be associated with poor quality information, resulting in stickiness akin to commitment \citep[e.g.,~][]{MM:04}.

\subsection{The Limit to Information Aggregation}

\label{sec:infoagg}

Our welfare conclusions for our Downsian model stem from the following result.

\begin{proposition}
\label{thm:welfare}\label{p:UD-best}Assume the signal distribution $F_{s_A,s_B}$ satisfies \autoref{completeness_general}, and consider any equilibrium with candidates' strategies $(y_A,y_B)$. There is a candidate $i \in \{A,B\}$ such that the voter's welfare in this equilibrium is the same as if candidate $i$ were elected no matter which policies are proposed using $(y_A,y_B)$.
\end{proposition}

To elaborate, consider any equilibrium $(y_{A},y_{B},w_{A})$. Denote by $v_{i}(y_{A},y_{B})$ the voter's welfare from electing candidate $i$ no matter which policies are proposed. Since the voter always has the option of electing one candidate regardless of the platforms, it holds that $v(y_A,y_B,w_A)\geq \max\{v_{A}(y_{A},y_{B}),v_B(y_A,y_B)\}$. \autoref{thm:welfare} says that under \autoref{completeness_general}, the bound is tight: 
\begin{equation*}
v(y_{A},y_{B},w_{A}) = \max \{v_{A}(y_{A},y_{B}),v_{B}(y_{A},y_{B})\}.
\end{equation*}
Put another way, insofar as equilibrium welfare is concerned, the voter may as well be ignoring one of the candidates and always electing the other. Crucially,
this is an ``as if'': both candidates may in fact win with positive probability in equilibrium, as detailed in \autoref{sec:pandering}.

\autoref{thm:welfare} is a straightforward application of \autoref{cor:singleoutcome}. Given an arbitrary voter strategy $w_A$ (recall $w_B\equiv 1-w_A)$, the two candidates are engaged in a constant-sum Bayesian game in which each candidate $i$'s payoff is the probability $w_{i}(x_{i},x_{-i})$ with which he wins the election. These payoffs are type independent because the candidates are office motivated and the voter can only infer their types from the platforms. As candidates have strict preferences over the winning probability outcome,  \autoref{cor:singleoutcome} implies that under \autoref{completeness_general}, in any equilibrium the probability of a candidate winning is independent of which on-path platforms are proposed. Hence, there are only two possibilities on the equilibrium path. Either (i) one candidate wins with probability one, or (ii) both candidates win with a constant interior probability, regardless of their platforms. In the latter case, the voter is always indifferent between the candidates. It follows that in either case, the voter's ex-ante expected utility can be evaluated as if she always elects the same candidate.

\autoref{thm:welfare} implies a sharp upper bound on the voter's welfare across all equilibria (or indeed, by the logic in the previous paragraph, given any voter strategy---including one she commits to ex ante---to which the candidates best respond). To make that precise, let $v^*_i$ denote the voter's welfare if candidate $i$ were always elected with his platform chosen, based on his information alone, to maximize voter welfare. \autoref%
{thm:welfare} implies that in any equilibrium, welfare is at most 
\begin{equation} 
\label{e:welfareupperbound}
\max\{v^*_A,v^*_B\}.
\end{equation}
In other words, even when both candidates have socially valuable information, the voter's equilibrium welfare is, at best, determined by the efficient use of only one candidate's signal.

\autoref{thm:competition} below states that point and also observes that the upper bound \eqref{e:welfareupperbound} can be achieved. Say that candidate $i$ is \emph{better} (than his opponent) if $v^*_i \geq v^*_{-i}$. That is, if each candidate would choose the voter-optimal policy based on their information alone, playing $y^*_i(s_i):=\arg \max_{x\in X}\mathbb{E}_{\theta}\left[u(x,\theta) \mid s_i\right]$,\footnote{If there are multiple maximizers at any signal, any measurable selection will do.} then the voter would prefer to ex-ante delegate policymaking to $i$ rather than the opponent. (If $v^*_A=v^*_B$, then without loss we stipulate that $A$ is the better candidate.)

\begin{theorem}
\label{thm:competition} \label{p:UD} If the signal distribution $F_{s_A,s_B}$ satisfies \autoref{completeness_general}, then a voter-welfare maximizing equilibrium has welfare $%
\max\{v^*_A,v^*_B\}$. There is one such equilibrium in which the better
candidate $i\in\{A,B\}$ is elected with probability one and plays $y^*_i$.
\end{theorem}

There may be multiple equilibria that achieve the proposition's welfare bound, but a simple construction is as follows. The better candidate $i$ plays $y^*_i$, and the opponent uninformatively chooses the prior-optimal policy, i.e., he plays $y_{-i}(s_{-i})=\arg \max_{x\in X}\mathbb{E}_{\theta }[u(x,\theta)]$. It is then optimal for the voter to always elect $i$ on path. We can stipulate that the voter also elects $i$ if  $-i$ chooses any other platform (and $i$ chooses any of his on-path platforms) because she believes that the deviation by $-i$ is uninformative about $s_{-i}$.\footnote{Any sequentially rational behavior by the voter after an observable deviation by $i$ supports the equilibrium, as $i$ has no incentive to deviate.}$^{,}$\footnote{\label{fn:50-50}Let $x_i$ be an on-path platform of candidate $i$. Our construction entails the voter electing candidate $i$ even if both candidates choose $x_i$. If one insists that the voter must randomize uniformly between the candidates when indifferent, then \autoref{thm:competition} is still valid with essentially the same construction so long as every on-path platform of candidate $i$ has zero ex-ante probability. This is the case with a continuous policy space when there is a unique and distinct optimal policy after each signal of the better candidate $i$ and the marginal distribution $F_{s_i}$ is atomless. An example is the normal-quadratic setting of \autoref{sec:pandering}.} Note that this construction does not require \autoref{completeness_general}; rather, the condition guarantees, by \autoref{thm:welfare}, that this equilibrium is welfare maximizing. In addition, while this construction uses off-path beliefs, the welfare bound does not hinge on them; \autoref{fn:no-off-path} in \autoref{sec:pandering} shows that in the normal-quadratic setting, the bound is also attained in an equilibrium in which every platform pair is on path. 

The following example  shows that \autoref{completeness_general} cannot be dispensed with in  \autoref{thm:competition}.

\begin{example}
    \label{eg:completeness-Downsian}
Let $X=S_A=S_B=\{1,2,3,4\}$, and $\Theta=S_A\times S_B$ with a uniform prior. In each state $\theta$, the signal profile is deterministically $(s_A,s_B)=\theta$. Hence, the unconditional joint signal distribution is uniform on $S_A\times S_B$, violating \autoref{completeness_general}. The voter's utility is  $1$ if the ``correct'' policy is implemented in a state and $0$ otherwise, with the correct policy in each state---or equivalently, after each signal profile $(s_A,s_B)$---shown in the left table below. The right table shows a strategy for the voter, i.e., which candidate she elects following each platform pair $(x_A,x_B)$.
\[
\begin{tabular}{cc}
\text{Correct policy} & \text{Elected candidate} \\[3pt]
\begin{tabular}{|c|c|c|c|c|}
\hline
$s_A \,/\, s_B$ & $1$ & $2$ & $3$ & $4$ \\
\hline
1 & 1 & 2 & 3 & 1 \\
\hline
2 & 1 & 2 & 3 & 2 \\
\hline
3 & 3 & 2 & 3 & 4 \\
\hline
4 & 1 & 4 & 4 & 4 \\
\hline
\end{tabular}
&
\begin{tabular}{|c|c|c|c|c|}
\hline
$x_A \,/\, x_B$ & $1$ & $2$ & $3$ & $4$ \\
\hline
1 & $A$ & $B$ & $B$ & $A$ \\
\hline
2 & $B$ & $A$ & $B$ & $A$ \\
\hline
3 & $A$ & $B$ & $A$ & $B$ \\
\hline
4 & $B$ & $A$ & $A$ & $B$ \\
\hline
\end{tabular}
\end{tabular}
\]

\noindent This voter strategy and each candidate playing the strategy $s_i\mapsto s_i$ constitute an equilibrium: the voter is playing optimally because she obtains her preferred policy in every state; candidates are playing optimally because, no matter their signal, their posterior is uniform over the opponent's signal and they thus expect to win with probability $1/2$ no matter their platform. This equilibrium achieves the voter's first-best welfare, which is larger than the welfare level $\max\{v^*_A,v^*_B\}$ because neither candidate's signal individually reveals the state. Evidently, the conclusions of \autoref{thm:welfare} and \autoref{thm:competition} do not hold.
\end{example}

\subsection{The Normal-Quadratic Specification}
\label{sec:pandering}

To substantiate  \autoref{thm:welfare} and \autoref{thm:competition}, we now elaborate on a leading specification of our Downsian model. The analysis of this subsection yields insights into how politicians' strategic incentives play out in equilibria---in particular, on whether office motivation necessarily leads to pandering, and whether pandering is detrimental to welfare.

Consider a one-dimensional \emph{normal-quadratic} specification: $X=\Theta=S_A=S_B=\mathbb{R}$, the voter's utility is $u(x,\theta )=-(x-\theta )^{2},$ the state is $\theta \sim \mathcal{N}(0,1/\alpha )$, and, conditional on the state $\theta$, each candidate $i\in \{A,B\}$ receives an independent signal $s_{i} \sim  \mathcal{N}(\theta,1/\beta )$, with parameters $\alpha,\beta \in \mathbb{R}_{>0}$. Note that the unconditional joint signal distribution satisfies \autoref{completeness_general}. \appendixref{app:exponential} discusses how some of this section's themes generalize to broader informational structures.

Quadratic-loss utility implies that the voter's preferred policy given any 
information $\mathcal{I}$ is $\mathbb{E}[\theta \mid \mathcal{I}]$. Hence, by
standard properties of normal information, 
\begin{equation}
y_{i}^{\ast }(s_{i})=\mathbb{E}\left[ \theta  \mid s_{i}\right] =\frac{\beta }{%
\alpha +\beta }s_{i},  \label{e:Bayes1}
\end{equation}%
and we refer to $y^*_i$ as the \emph{unbiased} strategy because it is the
best estimate of state given $s_i$. We say that a strategy $y_{i}$ displays 
\emph{pandering} (or underreaction) if 
$s_{i}> 0 \implies y_i(s_i)\in [0,\mathbb{E}[\theta \mid s_i])$, $s_{i}< 0
\implies y_i(s_i)\in (\mathbb{E}[\theta \mid s_i],0]$, and $y_i(0)=0$. 
In other words, a candidate panders if for $s_i\neq 0$ his platform is
distorted from his unbiased estimate toward the voter's prior expectation $%
\mathbb{E}[\theta ]=0$ of the best policy. Analogously, we say that $y_{i}$
displays \emph{anti-pandering} (or {overreaction}) if 
$s_{i}>0 \implies y_i(s_i)>\mathbb{E}[\theta \mid s_i]$ and $s_{i}< 0 \implies
y_i(s_i)<\mathbb{E}[\theta \mid s_i]$. We also say that a platform $x$ is \emph{%
more extreme} than platform $x^{\prime }$ if the former is further from the
prior mean of $0$, i.e., if $\abs{x}>\abs{x^{\prime }}$. A strategy $y_{i}$ is \emph{%
informative} if it is not constant, and it is \emph{fully revealing} if it
is injective. An equilibrium is \emph{symmetric} if both candidates use the same strategy
and both win with positive probability.

\paragraph{Unbiased Strategies.}
There are trivial equilibria in which candidates disregard their information, e.g., the ``full pandering equilibrium'' in which they each play  $y_i(\cdot)=\mathbb{
E}[\theta]=0$, and the voter elects  each candidate with some constant probability no matter the platforms. 
To tackle informative equilibria, a natural starting point is the profile of unbiased strategies.
From \autoref{e:Bayes1}, we see that the voter would then infer from a
platform $x_i$ that $s_i=\frac{\alpha+\beta}{\beta}x_i$. As the expected
value of $\theta $ conditional on both signals is 
\begin{equation}
\mathbb{E}[ \theta  \mid s_{A},s_{B}] =\frac{2\beta }{\alpha +2\beta }\left( 
\frac{s_{A}+s_{B}}{2}\right),  \label{e:Bayes2}
\end{equation}
the voter's posterior expectation of the state given the platforms $x_A$ and 
$x_B$ is 
\begin{equation*}
\frac{2(\alpha+\beta)}{\alpha+2\beta}\left(\frac{x_A+x_B}{2}\right).
\end{equation*}
So the voter's preferred policy, which is the posterior expectation, has the same sign as the average of the two platforms but is more extreme (so long as the average is non-zero). The voter thus elects the more extreme candidate, and consequently, each candidate would benefit by deviating to a more extreme platform, i.e., by anti-pandering or overreacting to his information. \autoref{thm:unbiased} in \appendixref{app:unbiased} provides a formal statement. 

\paragraph{Equilibrium Anti-Pandering.} Building on the above intuition, the next result identifies an
anti-pandering equilibrium.

\begin{proposition}
\label{thm:FRE}In the normal-quadratic specification, there is an anti-pandering equilibrium, which is symmetric and fully-revealing: both candidates play 
\begin{equation}
y_{i}(s_{i})=\mathbb{E}[\theta  \mid s_{i},s_{-i}=s_{i}]=\frac{2\beta }{\alpha
+2\beta }s_{i},  \label{e:FRE}
\end{equation}
and each candidate is elected with probability $1/2$ regardless of their
platforms.  The voter's welfare in this equilibrium is 
\begin{equation}
\label{e:FREwelfare}
-\frac{\alpha +4\beta }{\left( \alpha +2\beta \right)^{2}}.
\end{equation}
Moreover, any symmetric equilibrium in which both candidates use fully-revealing and continuous pure strategies has both candidates playing \eqref{e:FRE} and voter welfare \eqref{e:FREwelfare}.
\end{proposition}

In the equilibrium of \autoref{thm:FRE}, candidates can be viewed as choosing the unbiased platform based on a signal with twice the actual accuracy.
Alternatively, each candidate's platform is the Bayesian estimate of the
state assuming his opponent has received the same signal. That is despite each candidate $i$ knowing that, in expectation, his opponent's signal is in fact more moderate
than his own, as that expectation is just $i$'s unbiased estimate of the state, $\frac{\beta }{\alpha +\beta }s_{i}$. When
the voter conjectures that both candidates play the strategy \eqref{e:FRE},
she is indifferent between the candidates no matter their platforms. 
For, whenever a candidate $i$ increases his platform by any $\delta >0$, formula %
\eqref{e:Bayes2} implies that the voter's posterior expectation increases by 
$\frac{2\beta }{\alpha +2\beta }\left( \frac{\alpha +2\beta }{2\beta }\frac{%
\delta }{2}\right) =\delta /2$.

An important implication of \autoref{thm:FRE} is that office motivation does not necessarily lead to pandering. Moreover, the voter's welfare \eqref{e:FREwelfare} in the anti-pandering equilibrium is higher than in the trivial full-pandering equilibrium, as the welfare in the latter is simply $-1/\alpha$. Consistent with \autoref{thm:welfare}, 
both the anti-pandering equilibrium and the trivial full-pandering equilibrium have the ex-post property for the candidates, and the voter's welfare in %
these equilibria is the same as if she always elected either candidate. 
Moreover, consistent with the construction we described for \autoref{thm:competition}, there is yet another equilibrium: (either) candidate $i$ plays the unbiased strategy \eqref{e:Bayes1}, the other candidate $-i$ plays $y_{-i}(\cdot)=\E[\theta]=0$, and the voter always elects candidate $i$.\footnote{\label{fn:no-off-path}In this normal-quadratic specification, the same outcome---i.e., that $i$ plays \eqref{e:Bayes1} and is always elected---can also be supported in a fully-revealing equilibrium with $y_{-i}(s_{-i})=s_{-i}$. Every platform pair is then on path, so the voter's beliefs are fully determined by Bayes' rule. Candidate $-i$ is overreacting to his information here to such an extent that the voter never finds it optimal to elect $-i$ despite correctly inferring his information.} The voter's welfare in this equilibrium can be straightforwardly computed as $-1/(\alpha+\beta)$, which is even higher than the anti-pandering equilibrium's welfare \eqref{e:FREwelfare}. Indeed, although it is infeasible to characterize all equilibria of the normal-quadratic specification, \autoref{thm:competition} tells us that $-1/(\alpha+\beta)$ is the maximum equilibrium welfare.

\paragraph{The (Disequilibrium) Benefits of Pandering.}

Interestingly, in this normal-quadratic specification, an appropriate degree of non-equilibrium pandering would actually
benefit the voter. To get some intuition for why, consider again the benchmark where both politicians play the unbiased strategy $y^*_{i}(s_i)=\mathbb{E}[\theta  \mid s_{i}]$. %
As explained above, the voter would then select the politician with the most extreme platform. This implies a ``winner's curse'': the electoral winner, say $i$, would have received the most extreme signal, and so voter welfare would be improved if $i$ were elected with a slightly more moderate platform. Such moderation can be achieved by underreacting to private information, i.e., by pandering---although that runs counter to office motivation.

To formalize the point, consider the following strategy:
\begin{equation}
y_i(0)=0 \text{, and for $s_i\neq 0$, } y_i(s_{i})=\mathbb{E}\left[\theta \mid s_{i},\abs{s_{-i}}\leq \abs{s_{i}}\right],
\label{e:benevolent}
\end{equation}
which features pandering because conditioning on the opponent having a more
moderate signal makes a candidate underreact to his own signal. \autoref{lem:benevolent-br} in \appendixref{sec:panderingbenefit} verifies that, and also shows that the voter's best response to both candidates playing \eqref{e:benevolent} is to elect the candidate with the more extreme platform (which, recall, is also her best response to both candidates using unbiased strategies).

\begin{proposition}
\label{p:benevolent-best-weaker}
In the normal-quadratic specification, consider the strategy profile where both candidates pander by playing strategy \eqref{e:benevolent}, and the voter best responds. This profile yields higher voter welfare than any equilibrium, as well as the non-equilibrium profile in which the candidates play the unbiased strategies \eqref{e:Bayes1} and the voter best responds.
\end{proposition} 

In fact, we prove in \appendixref{sec:panderingbenefit} that the (non-equilibrium) strategy profile of \autoref{p:benevolent-best-weaker}  maximizes voter welfare within a broad class of profiles. Our takeaway is that an appropriate degree of pandering would benefit the voter. 

\subsection{Discussion}
\label{sec:elections-discussion}

We now return to our general Downsian model with informed candidates and discuss the robustness of our welfare conclusions.

\paragraph{Candidates Mixing.} \autoref{thm:welfare}, and hence the welfare bound of \autoref{thm:competition}, also apply to equilibria in which candidates mix. If at least one candidate's strategy is identifiable, then \autoref{thm:expost-general} delivers ex-postness, and the conclusion is immediate. Ex-postness can fail if neither candidate's strategy is identifiable, but \autoref{thm:welfare}'s welfare conclusion needs only that a candidate's win probability not vary with his own platform or his opponent's \emph{signal}, rather than his opponent's platform. For, a candidate winning is then independent of the state. We show in \appendixref{app:mixing} that this constancy obtains under \autoref{completeness_general}, building on \autoref{lem:zerosum_general}.

To illustrate, consider a variant of \autoref{eg:completeness}: $\Theta =X=\{1,2\}$, $u(x,\theta )=\ind\{x=\theta\}$, a uniform prior on $\Theta $, and any signal structure $F_{s_A,s_B}$ that satisfies \autoref{completeness_general}. There is an equilibrium in which, regardless of their signals, both candidates mix uniformly over both policies, and the voter (being indifferent between both policies) plays $w_{A}(x_{A},x_{B})=\ind\{x_{A}=x_{B}\}$. Neither candidate's strategy is identifiable, and the equilibrium is not ex post: a candidate's win probability is either $1$ or $0$ depending on the platform pair. Yet conditional on his opponent's signal, each candidate wins with probability $1/2$ whichever policy he proposes, and the voter's welfare is the same as if she always elected one of them, as in \autoref{thm:welfare}.

\paragraph{Other Game Forms.}
\autoref{thm:welfare} also applies much more generally than to the canonical  Downsian game form we have considered. For concreteness, we only mention two variations:
\begin{enumerate}
    \item \label{noisy-implementation} The elected candidate does not necessarily implement their platform $x_i$, but instead some exogenous---possibly stochastic---function of $x_i$. For example, there could be a status quo policy $x^0$ (e.g., the ex-ante optimal policy), and the elected candidate $i$ implements their platform $x_i$ with some probability (which could depend on $x_i$) and $x^0$ otherwise. Alternatively, the candidate may always moderate after the election and implement $q_i\cdot x_i$ for some parameter $q_i\in(0,1)$.
    \item \label{sequential} Instead of choosing platforms simultaneously, candidate $A$ chooses his platform $x_A$ first, and candidate $B$, having observed $x_A$, then chooses $x_B$.
    The asymmetry in timing might reflect that one candidate is an incumbent and the other a challenger.
\end{enumerate}
The reason \autoref{thm:welfare} holds for these variations is that the candidates are still purely office motivated and the voter's decision cannot depend directly on their signals; hence, given any voter strategy, the candidates still face a constant-sum Bayesian game with type-independent payoffs, and \autoref{thm:expost-general} applies.

On the other hand, the welfare level obtained in \autoref{thm:competition} may not apply to these variations. In model variation \#\ref{noisy-implementation} above, the upper bound on equilibrium welfare would have to be adjusted for the stochastic policy implementation; if the status quo is implemented with high probability, then evidently equilibrium voter welfare cannot be much higher than the expected utility from the status quo. More importantly, model variation \#\ref{sequential} above generally allows for equilibria that achieve welfare higher than $\max\{v^*_A,v^*_B\}$. For, the level $v^*_i$ obtains from efficiently using only candidate $i$'s signal. Under natural specifications, there can even be full information aggregation in model variation \#\ref{sequential}: candidate $A$ reveals his signal $s_A$ via his platform; candidate $B$ then proposes the best policy for the voter given both $s_A$ and $s_B$; and the voter always selects candidate $B$. Certainly this raises questions about equilibrium refinements.\footnote{Indeed, there could be another equilibrium in which candidate $A$---without loss, the better candidate---proposes the best policy given his signal $s_A$ (which reveals his signal); candidate $B$ then proposes the \emph{worst} policy for the voter given both candidates' signals; and voter always selects candidate $A$. This equilibrium's welfare is the same as that of \autoref{thm:competition}.} We do not pursue that issue; instead, we observe that the welfare bound of \autoref{thm:competition} %
is focal because it applies to the canonical Downsian game form.

\paragraph{Voter Commitment.} For some interpretations of the model---such as decision-making in an organization, as mentioned earlier---it is plausible that the voter (decision maker) can commit ex ante to how she will select among the candidates' (agents') platforms (proposals). Since the constant-sum property between candidates holds for an arbitrary voter strategy, \autoref{thm:welfare} and \autoref{thm:competition} also apply to this case. In other words, commitment cannot increase the maximum voter welfare.

\paragraph{Beyond Office Motivation.}

Since candidates' office motivation is a key assumption for our results, we
conclude this subsection by discussing the robustness of \autoref{thm:competition}'s welfare conclusion to small departures from that
assumption. 

Consider a variant of our Downsian model in which the payoff of each
candidate $i\in \{A,B\}$ is given by $u_{i}(x_{A},x_{B},\theta ,W;\gamma
_{i}),$ where the new notation $W\in \{A,B\}$ denotes the election's winner
and $\gamma _{i}$ is a commonly-known payoff parameter. Pure
office-motivation corresponds to the utility 
$\mathbbm{1}{\{W=i\}}$, but in general $u_{i}(\cdot )$ allows for a variety
of \emph{mixed motivations}, including policy motivation (a candidate cares
about the winner's policy, in relation to the state) and platform motivation
(he cares about his own platform, in relation to the state).

An election with mixed motivations is not generally a constant-sum game for the candidates; consequently, for arbitrary mixed motivations, voter welfare may be significantly different from the bound in \autoref{thm:competition}. However, consider a family of mixed-motivations games in which each candidate $i$'s payoff is parameterized by $\gamma_i\in \mathbb{R}^m$ such that $u_i(x_A,x_B,\theta,W;\vec{0})=\mathbbm{1}\{W=i\}$. That is, when $\gamma_i=\vec{0} \equiv (0,\ldots,0)$, candidate $i$ is purely office motivated. Under appropriate technical conditions, the equilibrium correspondence is upper hemicontinuous in the parameter $(\gamma_A,\gamma_B)$, and hence the upper bound on voter welfare when $(\gamma_A,\gamma_B)\approx (\vec{0},\vec{0})$, i.e., when candidates are almost office-motivated, is approximately that of \autoref{thm:competition}.%
\footnote{More precisely, we would be assured upper hemicontinuity of the set of Bayes-Nash equilibria. \autoref{thm:welfare} holds for Bayes-Nash equilibria too, because it does not use voter optimality off the equilibrium path; mixing by candidates is covered by \autoref{p:mixing} in \appendixref{app:mixing}.}
Simple sufficient technical conditions are that all the spaces $S_A$, $S_B$, $\Theta$, and $X$ are finite and that each $u_i(\cdot)$ is continuous in %
$\gamma_i$.

We note that our leading one-dimensional normal-quadratic specification from \autoref{sec:pandering} does not satisfy the
aforementioned technical conditions; in particular, the policy space $X=%
\mathbb{R}$ is not compact. The \hyperref[app:mixed]{Supplementary Appendix} analyzes an extension of
the normal-quadratic model with mixed motivations of the form 
\begin{equation}
u_{i}(x,\theta ,W;b_{i},\rho _{i})=-\rho _{i}(x_{W}-\theta
-b_{i})^{2}+(1-\rho _{i})\mathbbm{1}{\{W=i\}}.  \label{e:mixed-motivations}
\end{equation}%
So each candidate $i$ has quadratic-loss policy utility with an ideological
bias $b_{i}\in \mathbb{R}$ and places weight $\rho _{i}\in [0,1]$ on
policy utility. %
For this specification, we establish that among equilibria in monotone strategies satisfying a linear growth bound, the upper bound on voter welfare when each $b_{i}\approx 0$ and $\rho _{i}\approx 0$ is still close to that of efficiently using only one candidate's signal. Moreover, there is an equilibrium in the class that approximately achieves that welfare. So, the welfare conclusions of \autoref{p:UD} still approximately hold, at least in this equilibrium class.

\section{Competition in Dual Spheres}
\label{sec:dualcompetition}

To illustrate the implications of \autoref{thm:expost-general} beyond elections, we now develop an application involving competition in dual spheres. We frame it as two firms competing to both maximize their shares of a private market and to secure a government allocation. So, as in \autoref{sec:main}, we have two agents (politicians previously, now firms) competing for the favor of a principal (an electorate previously, now a government). However, the agents' actions now  affect their own payoffs directly, rather than only through the principal's response. This changes what the ex-post property delivers: instead of pinning down the principal's strategy, it ties her actions to the agents' direct payoffs. At the end of this section we discuss how the framework is more broadly applicable; indeed, it formally subsumes our Downsian election model. 

Consider two firms, $A$ and $B$. Each firm $i \in \{A,B\}$ chooses a product or technology $x_i \in X_i$. There is an unknown state of the world $\theta \in \Theta\subset \Reals^n$, representing a variable that matters for a government's action (elaborated below); for instance, $\theta$ could reflect the relative social value of the firms. Each firm $i$ observes a private signal $s_i \in S_i\subset \Reals^n$ about the state, drawn from some joint distribution conditional on the state $F_{s_A,s_B \mid \theta}$. Denote the unconditional joint distribution of signals by $F_{s_A,s_B}$.\footnote{We suppress the technical conditions on $\Theta$, each $S_i$, and the distributions, which follow those in the previous sections; in particular, $F_{s_A,s_B}$ is absolutely continuous with respect to the product of its marginals.}

The firms’ choices have consequences in two domains. First, they determine market shares in a private market, captured by a function $\market: X_A \times X_B \to [0,1]$, where $\market(x_A, x_B)$ is firm $A$'s share. This private-market competition is constant sum: it contributes a payoff $\market(x_A, x_B)$ to firm $A$ and $1 - \market(x_A, x_B)$ to firm $B$. Note that the function $\market$ does not depend on the state $\theta$; the state represents social value rather than appeal in the private market. This is reasonable when $\theta$ represents externalities, long-run reliability, or other attributes that the market does not price.

Second, a government observes some statistic $t\in \mathcal T$ of the firms' choices, generated by the map $\tau(x_A,x_B)$, and then chooses an action or allocation $a\in \mathcal{A}$. The government's payoff is given by $u_G(x_A,x_B,a,\theta)$. For instance, $a$ may represent the share of public procurement allocated to firm $A$, 
and the payoff $u_G$ may represent how well this allocation matches the state, with higher values 
of $\theta$ leading the government to prefer larger shares for firm $A$. That could be captured by $a\in [0,1]$, $\theta \in \Reals$, and $u_G = -(a-\theta)^2$.

The firms care about both their private market share and the government action (e.g., public procurement share). For some bounded function $v:\mathcal A\to \Reals$, firm $A$'s overall payoff is
\[
u_A(x_A, x_B, a) := \market(x_A, x_B) + v(a),
\]
and firm $B$'s, after normalization, is
\[
u_B(x_A, x_B, a) := - \market(x_A, x_B) - v(a).
\]

We refer to this setting as one of \emph{competition in dual spheres}, because the firms are competing for both market share and the government action. 

Observe that any government strategy $\alpha:\mathcal{T} \to \Delta(\mathcal A)$ induces a constant-sum Bayesian game with type-independent payoffs between the firms in which $A$'s payoff is $\market(x_A,x_B)+v_\alpha(x_A,x_B)$, where we define $v_\alpha:(x_A,x_B)\mapsto \E\left[v(\alpha(\tau(x_A,x_B)))\right]$, with the expectation over the government's randomization.\footnote{We assume this expectation is well-defined; we suppress such technical details in the rest of this section.} Denoting firm $i$'s strategy by $\sigma_i:S_i\to \Delta(X_i)$, the following result follows immediately from %
\autoref{cor:singleoutcome}.

\begin{proposition}
\label{p:dualcompetition}
Consider competition in dual spheres, with $F_{s_A, s_B}$ satisfying \autoref{completeness_general}. In any Bayes-Nash equilibrium $(\sigma^*_A,\sigma^*_B,\alpha^*)$ in which firms play identifiable strategies, %
$\market(x_A, x_B) + v_{\alpha^*}(x_A,x_B) $ is a.s.~constant on path.
\end{proposition}

The proposition establishes that---under \autoref{completeness_general} and identifiability, qualifiers we omit in the subsequent paragraphs for brevity---the firms' dual competition forces a rigid relationship between the private market and the government action: whatever the government's objective, the firms' expected valuation of its action must perfectly offset the market outcome. Although stated for (Bayes-Nash) equilibria of the game, the result evidently holds for any government strategy to which the firms mutually best respond. 

\autoref{p:dualcompetition} implies strong constraints on information revelation and government welfare. We illustrate with the following corollary, which assumes the government observes at least the market outcome, i.e., $\market$ is measurable with respect to $\tau$; the leading case is $\tau(\cdot)=\market(\cdot)$.

\begin{corollary}
\label{cor:crowdingout}
Under the hypotheses of \autoref{p:dualcompetition} and when the government observes at least the market outcome, it holds in any equilibrium that:
\begin{enumerate}
\item \label{crowdingout1} If the firms do not value the government's action ($v(\cdot)=0$), then the market outcome is a.s.~constant on path.
\item \label{crowdingout2} Suppose $\mathcal A=[0,1]$, $v(a)=a$, and $u_G(a,\theta)=-(a-\theta)^2$, where $\theta\in[0,1]$. Then $\Cov(\market,\theta)=-\Var(\market)\leq 0$.
\end{enumerate}
\end{corollary}

Part \ref{crowdingout1} of the corollary is immediate from \autoref{p:dualcompetition}. In part \ref{crowdingout2}, the government's optimal action is its posterior mean of the state, so $v_{\alpha^*}=\E[\theta\mid\tau]$. Since the market outcome $\market$ is $\tau$-measurable, the law of iterated expectations gives $\Cov(\market,\theta)=\Cov(\market,\E[\theta\mid\tau])$, and since \autoref{p:dualcompetition} implies $\E[\theta\mid\tau]$ differs from $-\market$ by a constant, this equals $\Cov(\market,-\market)=-\Var(\market)$.

In part \ref{crowdingout1}, the market outcome carries no information; if the government observes nothing else, its allocation cannot vary with the fundamental. Part \ref{crowdingout2} does not preclude an informative market outcome, but then it must be negatively correlated with the fundamental. What equilibrium rules out is the natural configuration in which greater market success is evidence of higher social value. The key culprit is the conjunction of the firms' private-market competition and their valuing the government's action in opposing ways.\footnote{\label{fn:dualcompetition-countereg}The following variation of \autoref{eg:completeness} shows how both parts of the corollary can fail without \autoref{completeness_general}. Let the distribution on $S_A\times S_B=\{0,1\}\times \{0,1\}$ be uniform (violating \autoref{completeness_general}), let $X_A=X_B=\{0,1\}$ and $\market(x_A,x_B)=\abs{x_A-x_B}$, and take $\theta=\ind\{s_A\neq s_B\}$. Suppose each firm plays the pure strategy $s_i\mapsto s_i$. From each firm's perspective, taking the other firm's strategy as given, either of its own actions induces a uniform lottery over market outcomes $0$ and $1$, so the firms are indifferent and we have an equilibrium with identifiable strategies. The market outcome fully reveals $\theta$ both when (i) $v(\cdot)=0$ and when (ii) $v(a)=a$ and the government optimally chooses $a=\E[\theta\mid \market]=\market$, with $\Cov(\market,\theta)=\Var(\market)>0$ in this case.}

\begin{remark}
\label{rem:dualcompetition}
The framework of this section is in some respects quite general and relevant to various economic contexts, as we now illustrate:
\begin{enumerate}
\item It subsumes the Downsian election model from \autoref{sec:main}. Concretely, the Downsian one-dimensional quadratic specification obtains when $X_A=X_B=\Theta=\Reals$, $\tau(x_A,x_B)=(x_A,x_B)$, $\mathcal A=\{A,B\}$, $u_G(x_A,x_B,a,\theta)=-(x_a-\theta)^2$, and $\market(\cdot)=0$ and \mbox{$v(a)=\ind\{a=A\}$}. This corresponds to the principal (voter) selecting a winning agent (candidate) to minimize the distance between the winner's action and the state,
while the agents just want to be selected.
\item %
Consider a modification of the above specification to $u_G(a,\theta)=-\left(\ind\{a=A\}-\theta\right)^2$. Then the agents' actions are cheap-talk messages, and the principal wants to select agent $A$ when the state is high. This is now a model of arbitration with a binary allocation, similar to \citet{Kattwinkel22}. \autoref{p:dualcompetition} implies that under its conditions, the principal's (probabilistic) selection in any equilibrium---or, even in a stochastic mechanism that the principal commits to, as the result holds for any principal strategy---must be independent of the agents' signals, as those authors also show for the case of finite signals. 

But \autoref{p:dualcompetition} can be applied to richer arbitration problems as well. Consider $\mathcal A=[0,1]$, where $a$ represents the arbitrator's ruling of a transfer from agent $B$ to $A$, and any strictly increasing affine $v(a)$ and any $u_G(a,\theta)$. Assume the arbitrator has a unique prior-optimal ruling, $a^*:=\arg\max_{a\in \mathcal{A}} \E_{\theta}\left[{u_G(a,\theta)}\right]$. %
Then \autoref{p:dualcompetition} implies that under its conditions, the \emph{expected} transfer in any equilibrium is constant across on-path signal profiles. If $u_G(\cdot,\theta)$ is in addition strictly concave for each $\theta$, then the arbitrator's posterior-optimal ruling is unique; the realized transfer is thus constant, and being optimal after every on-path profile, it is optimal ex ante and so equals $a^*$.

\item The principal can be passive while caring about the agents' interaction and the state.  For instance, agent $B$ may be a regulator (or security force) chasing a non-compliant firm $A$ (or interdicting a smuggler). Each agent $i$ chooses a location $x_i \in X_i\subset \Reals^k$ and they have opposing preferences over their location gap or compliance: $\market(x_A, x_B) = \norm{x_A - x_B}$ for some norm $\norm{\cdot}$ with values normalized to lie in $[0,1]$, and $v(\cdot)=0$. The principal/society takes no action ($\abs{\mathcal {A}}=1$) but has utility $u_G(x_A,x_B,\theta)= -\theta \norm{x_A - x_B}$. So society would prefer tighter compliance when the stakes $\theta\in \Theta \subset \Reals_{\geq 0}$ are higher. Although the agents are informed about these stakes, \autoref{p:dualcompetition} implies that under its conditions, in any equilibrium the realized compliance does not vary with signal profiles or stakes.\footnote{Even though the agents do not care about the stakes, absent \autoref{completeness_general} there are examples in which society gets its first-best state-contingent outcome; cf. \autoref{fn:dualcompetition-countereg}.}
\end{enumerate}
\end{remark}

\section{Conclusion}

We have developed a general result about equilibrium behavior in two-player constant-sum Bayesian games with type-independent payoffs. Under a completeness statistical condition on the distribution of types,  any identifiable (Bayes-Nash) equilibrium must be ex post: each player is indifferent among all the actions played by any of his types, even after observing the opponent’s action. This ex-postness property limits the extent to which adversarial players' information can be revealed to and used by third parties.

Motivated by the debate on whether political competition promotes information aggregation and informed choices by electorates, we have applied the above result to Downsian electoral competition between two office-motivated candidates who have private information about policy consequences. Under the completeness condition, we find a sharp bound on the (median or representative) voter's welfare. Welfare in any equilibrium is effectively determined by just one candidate's platform strategy. Consequently, Downsian elections cannot efficiently aggregate more than one candidate's information, despite the availability of two informational sources. Moreover, the upper bound of efficiently aggregating the ``better'' candidate's information can be achieved in an equilibrium. 

To substantiate the electoral welfare bound and to better  understand politicians' strategic incentives, we have studied in more detail a  normal-quadratic specification of our Downsian model. In that specification, there is a fully-revealing equilibrium in which candidates anti-pander or overreact to their information. Furthermore, we find that an appropriate degree of (disequilibrium) pandering by candidates would actually benefit voters. These findings run counter to conventional views that candidates' pandering is an inevitable consequence of candidate office-motivation, and is necessarily harmful to voters.

As in most formal models of spatial electoral competition, we have restricted attention to two candidates and assumed that their information is exogenously given. Relaxing both these assumptions are interesting topics for future research. We note here that since a voter-optimal equilibrium of our model involves always electing the ``better''---roughly, more informed---candidate, there can be strong incentives for candidates to observably acquire information.

While our primary application is to electoral competition, the logic of \autoref{thm:expost-general} applies more broadly. We have illustrated this with a model of competition in dual spheres, in which firms compete for both private market share and government actions. Even though the firms may be well informed about a fundamental that the government cares about, their constant-sum rivalry forces a rigid relationship between market outcomes and government actions---for instance, it can preclude state-contingent government procurement. This application underscores that the limits to information revelation and aggregation identified in this paper are not specific to elections, but are a general feature of institutions constrained by purely adversarial incentives.

\appendix

\section{Proofs and Other Material for \autoref{sec:main}}

\label{app:proofs} %

We omit proofs for \autoref{cor:singleoutcome}, \autoref{thm:welfare}, and \autoref{p:UD} (and also \autoref{p:dualcompetition} and \autoref{cor:crowdingout} in \autoref{sec:dualcompetition}), as they were explained in the main text.

\subsection{Unbiased Strategies}
\label{app:unbiased}

Let us substantiate the discussion in \autoref{sec:pandering} by showing that candidates cannot play unbiased strategies in an equilibrium of the normal-quadratic specification.

\begin{proposition}
\label{thm:unbiased} \label{uno} In the normal-quadratic specification, the profile of unbiased strategies cannot
be supported in an equilibrium. In particular, candidates would deviate by
overreacting to their information, whereas underreacting would be worse than
playing the unbiased strategy.
\end{proposition}

\begin{proof}
\hypertarget{proof:uno}{A}ssume both candidates use the unbiased strategy $%
y_{i}(s_{i})=\frac{\beta }{\alpha +\beta }s_{i}$. 
Since this strategy is fully revealing, the voter correctly infers $s_{A},s_{B}$ for all signal realizations. The voter's expected utility from a platform $x$ follows a standard mean-variance decomposition: 
\begin{align}
\mathbb{E}[u( x,\theta )  \mid s_{A},s_{B}]& =-\mathbb{E}\left[ \left( x-\theta
\right) ^{2} \mid s_{A},s_{B}\right]  \notag \\
& =-\left[ x^{2}+\mathbb{E}\left[\theta ^{2} \mid s_{A},s_{B}\right]-2x\mathbb{E}\left[\theta
 \mid s_{A},s_{B}\right]\right]  \notag \\
& =-\left[ x^{2}+\left(\mathbb{E}\left[\theta  \mid s_{A},s_{B}\right]\right)^{2}-2x\mathbb{E}\left[\theta
 \mid s_{A},s_{B}\right]\right] -\mathbb{E}\left[\theta ^{2} \mid s_{A},s_{B}\right]+\left(\mathbb{E}\left[\theta
 \mid s_{A},s_{B}\right]\right)^{2}  \notag \\
& =-\left[ x-\mathbb{E}(\theta  \mid s_{A},s_{B})\right] ^{2}-\Var\left( \theta
 \mid s_{A},s_{B}\right) .  \label{me-va}
\end{align}%
So the voter elects candidate $i$ whenever $x_{i}$ is closer to $\mathbb{E}[\theta  \mid s_{A},s_{B}]$ than is $x_{-i}$.

We now show that for any $i=A,B$ and $s_{i}$, candidate $i$ can profitably deviate. By \eqref{me-va}, if $i$ plays as if he has received signal $\hat s_i$ (no matter his true signal), then $i$ wins against any realization $s_{-i}$ such that 
\begin{equation*}
\left( y_{-i}\left( s_{-i}\right) -\mathbb{E}\left[ \theta  \mid \hat s_{i},s_{-i}
\right] \right) ^{2}>\left( y_{i}(\hat s_{i}) -\mathbb{E}\left[ \theta
 \mid \hat s_{i},s_{-i}\right] \right) ^{2}.
\end{equation*}%
Substituting from \eqref{e:Bayes1} and \eqref{e:Bayes2}, this is equivalent to 
\begin{equation*}
\left( \frac{\beta }{\alpha +\beta }s_{-i}-\frac{\beta }{\alpha +2\beta }(\hat s_{i}+s_{-i})\right) ^{2}>\left( \frac{\beta }{\alpha +\beta }\hat s_{i}-\frac{\beta }{\alpha +2\beta }(\hat s_{i}+s_{-i})\right) ^{2},
\end{equation*}
or after algebraic simplification, $(\hat s_{i})^{2}>(s_{-i})^{2}$. Hence, conditional on any signal $s_i$, candidate $i$'s winning probability after mimicking $\hat s_i$ is $\Pr\!\left(\abs{s_{-i}}<\abs{\hat s_i}\mid s_i\right)$. Because $s_{-i}\mid s_i$ is normally distributed, this probability is strictly increasing in $\abs{\hat s_i}$. Thus candidate $i$ strictly raises his winning probability by overreacting and strictly lowers it by underreacting.
\end{proof}

\subsection{Anti-Pandering}

\begin{proof}[Proof of \autoref{thm:FRE}]
For the proposition's first statement, it suffices to verify that the voter
is indifferent between the two candidates for any pair of platforms,
assuming that both candidates play the strategy \eqref{e:FRE}. Since the
candidates' strategies are fully revealing, the voter correctly infers the
candidates' signals from the platform pair. Furthermore, since the
candidates' strategies each have range $\mathbb{R}$, there are no off-path platform
pairs. Therefore, it suffices to show that for any $s_{i}$ and $s_{-i}$, we have $$-%
\mathbb{E}[(y_{i}(s_{i})-\theta )^{2} \mid s_{i},s_{-i}]=-\mathbb{E}[(y_{-i}(s_{-i})-\theta )^{2} \mid s_{i},s_{-i}],$$ or equivalently that $\left(
y_{i}(s_{i}) -\mathbb{E}\left[ \theta \mid s_{i},s_{-i}\right] \right)
^{2}=\left( y_{-i}\left( s_{-i}\right) -\mathbb{E}\left[ \theta \mid s_{i},s_{-i}
\right] \right) ^{2}$.\footnote{That this latter equality is equivalent to the former follows from a
standard mean-variance decomposition under quadratic loss utility as in the \hyperlink{proof:uno}{proof of \autoref{uno}}.} Using \eqref{e:Bayes2} and \eqref{e:FRE}, this latter equality can be rewritten as 
\begin{equation*}
\left( \frac{2\beta }{\alpha +2\beta }s_{i}-\frac{2\beta }{\alpha +2\beta }%
\left( \frac{s_{i}+s_{-i}}{2}\right) \right) ^{2}=\left( \frac{2\beta }{%
\alpha +2\beta }s_{-i}-\frac{2\beta }{\alpha +2\beta }\left( \frac{%
s_{i}+s_{-i}}{2}\right) \right) ^{2},
\end{equation*}%
which holds for any $s_{i}$, $s_{-i}$.

Next, we derive the equilibrium welfare \eqref{e:FREwelfare} as follows:
\begin{align*}
\E[-(y_i - \theta)^2]
&= \E\left[-\left( \frac{2\beta}{\alpha + 2\beta}s_i - \theta \right)^2\right] \\
&= -\Var\left( \frac{2\beta}{\alpha + 2\beta}s_i - \theta \right) \\
&= -\left( \frac{2\beta}{\alpha + 2\beta} \right)^2 \Var(s_i) - \Var(\theta) + 2 \left(\frac{2\beta}{\alpha + 2\beta}\right) \Cov(s_i, \theta) \\
&= -\left( \frac{2\beta}{\alpha + 2\beta} \right)^2 \left( \frac{1}{\alpha} + \frac{1}{\beta} \right) - \frac{1}{\alpha} + 2 \left( \frac{2\beta}{\alpha + 2\beta}\right)\frac{1}{\alpha} \\
&= -\frac{\alpha + 4\beta}{(\alpha + 2\beta)^2}.
\end{align*}

Finally, for the proposition's last statement, 
we prove something stronger that does not assume symmetry: in any equilibrium in which both candidates
win with positive probability and use continuous fully-revealing pure strategies,
there is $c\in \mathbb{R}$ and $i\in \{A,B\}$ such that 
\begin{equation*}
y_{i}(s_{i})=\frac{2\beta }{\alpha +2\beta }s_{i}+c \quad \text{and} \quad %
y_{-i}(s_{-i})=\frac{2\beta }{\alpha +2\beta }s_{-i}-c.\footnote{Using a very similar analysis to that in the first paragraph of this proof, it is readily verified that these strategies constitute an equilibrium, with the voter indifferent after any pair of platforms. %
Voter welfare is then $c^{2}$ lower than \eqref{e:FREwelfare}.
}
\end{equation*}
(Imposing symmetry, as in the proposition, implies $c=0$, which yields \autoref{e:FRE}.)  To prove that, fix any equilibrium in which each candidate $i$ uses a continuous and fully revealing strategy $\bar{y}_i$ and both win with positive probability. Denote the interior of the range of $\bar{y}_i $ by $\bar{X}_i$, noting that $\bar{X}_i$ is an open interval. Also denote $\bar{s}_i(x_i):=(\bar{y}_i)^{-1}(x_i)$. \autoref{thm:expost-general} and voter optimality imply that the voter is indifferent between both candidates after almost all on-path platform pairs. This implies that for almost all $x^{\prime }_{A}\in \bar{X}_A$ and $x^{\prime }_B\in \bar{X}_B$, we must have $\mathbb{E}[\theta \mid x^{\prime }_{A},x^{\prime }_{B}]=({x^{\prime }_{A}+x^{\prime }_{B}})/{2}$. Both sides of the equality are continuous, and the platform pair's distribution has full support on $\bar{X}_A\times \bar{X}_B$, so the equality holds for all $x^{\prime }_{A}\in \bar{X}_A$ and $x^{\prime }_B\in \bar{X}_B$. It implies \mbox{$\frac{\beta}{\alpha + 2\beta}(\bar{s}_A(x'_A)+\bar{s}_B(x'_B))=\frac{x'_A + x'_B}{2}$}, or equivalently 
\begin{equation}
\bar{s}_B(x^{\prime }_{B})=\frac{\alpha +2\beta }{2\beta }\left( x^{\prime
}_{A}+x^{\prime }_{B}\right) -\bar{s}_A(x^{\prime }_{A}).
\label{e:FRE-unique-1}
\end{equation}
For small $\epsilon>0$ and $x_A\in \bar{X}_A$ and $x_B\in \bar{X}_B$, the same logic also holds for platforms $x_A+\epsilon$ and $x_B-\epsilon$, yielding 
\begin{equation}
\bar{s}_{B}(x_{B}-\epsilon )=\frac{\alpha +2\beta }{2\beta }\left(x_{A}+x_{B}\right) -\bar{s}_{A}(x_{A}+\epsilon ).  \label{e:FRE-unique-2}
\end{equation}%
Substituting $x_{B}^{\prime }=x_{B}-\epsilon $ and $x_{A}^{\prime }=x_{A}$ into \eqref{e:FRE-unique-1} and then equating that with \eqref{e:FRE-unique-2} yields 
\begin{equation*}
\frac{\alpha +2\beta }{2\beta }\left( x_{A}+x_{B}-\epsilon \right) -\bar{s}_{A}(x_{A})=\frac{\alpha +2\beta }{2\beta }\left( x_{A}+x_{B}\right) -\bar{s}_{A}(x_{A}+\epsilon ),
\end{equation*}
or equivalently, 
\begin{equation}
\bar{s}_{A}(x_{A}+\epsilon )=\frac{\alpha +2\beta }{2\beta }\epsilon +\bar{s}_{A}(x_{A}).  \label{e:FRE-unique-3}
\end{equation}
The equality in \eqref{e:FRE-unique-3} can only hold for all $x_A\in \bar{X}_A$ and small $\epsilon>0$ if there is a constant $c_{A}\in \mathbb{R}$ such that $\bar{s}_{A}(x_{A})=\frac{\alpha +2\beta }{2\beta }\left( x_{A}-c_{A}\right) $ for all $x_{A}\in \bar{X}_A$, from which it follows that \mbox{$\bar{y}_A(s_A)=\frac{2\beta}{\alpha+2\beta}s_A+c_A$} for all $s_A$.
A symmetric argument establishes that \mbox{$\bar{y}_B(s_B)=\frac{2\beta}{\alpha+2\beta}s_B+c_B$} for all $s_B$. But then \eqref{e:FRE-unique-1} implies $c_{B}=-c_{A}$. Finally, note that continuity pins down the strategies even at measure zero sets of signals.
\end{proof}

\subsection{The Benefits of Pandering}
\label{sec:panderingbenefit}

\autoref{p:benevolent-best} below provides a result that is stronger than \autoref{p:benevolent-best-weaker}. Before that, we record the following.

\begin{lemma}
\label{lem:benevolent-br}
Assume the normal-quadratic specification. The strategy \eqref{e:benevolent} exhibits pandering. Moreover, if both candidates play \eqref{e:benevolent}, then the voter's best response is to elect the candidate with the more extreme platform.
\end{lemma}

\begin{proof}
We first prove that the strategy \eqref{e:benevolent} exhibits pandering. Using \eqref{e:Bayes2} and iterated expectations, and dropping the subscript on $y_i$ for the remainder of this proof (since the strategy is common to both candidates), we can rewrite 
$\eqref{e:benevolent}$
\begin{align}
y(s_i)
& =\E\big[\E[\theta \mid s_i, s_{-i}] \mid s_i,\abs{s_{-i}} \leq\abs{s_i}\big] \notag \\
& =\frac{\beta}{\alpha+2\beta}\Big(s_i+\E\big[s_{-i} \mid s_i,\abs{s_{-i}} \leq \abs{s_i}\big]\Big) \label{e:benevolent-rewrite}
\end{align} 
for $s_i\neq 0$, with $y(0)=0$. We will argue that if $s_i>0$ then $y(s_i)\in \bigl(0,\E[\theta \mid s_i]\bigr)$. By a symmetric argument for $s_i<0$, it follows that $y(\cdot)$ exhibits pandering.

Accordingly, fix any $s_i>0$. It is straightforward from \eqref{e:benevolent-rewrite} that $y(s_i)>0$. Next, algebraic manipulation of \eqref{e:benevolent-rewrite} and the equalities $\E[\theta \mid s_i]=\E[s_{-i} \mid s_i]=\frac{\beta}{\alpha+\beta}s_i$ shows that \mbox{$y(s_i)<\E[\theta \mid s_i]$} is equivalent to 
$$\E\big[s_{-i} \mid s_i,\abs{s_{-i}} \leq \abs{s_i}\big]<\E[s_{-i} \mid s_i].$$
This inequality holds because $s_{-i} \mid s_i$ is normally distributed with a mean $\frac{\beta}{\alpha+\beta}s_i>0$, and a truncation to the interval $\left[-s_i, s_i\right]$ which is symmetric around 0 (hence centered below the mean) pulls the truncated mean towards $0$.

Now we turn to the voter's best response.  Define the function $l:\Reals \to \Reals$ by
\[
l(0):=0 \text{, and for $s_i\neq 0$, } l(s_i):= s_i-\E\big[s_{-i} \mid s_i,\ \abs{s_{-i}}\le \abs{s_i}\big].
\]
Using \eqref{e:Bayes2} again and the formulae above, some algebra yields
\begin{equation}
\label{e:benevolent-br-1}
\E[\theta \mid s_A,s_B]-\frac{y_A(s_A)+y_B(s_B)}{2}
=\frac{\beta}{2(\alpha+2\beta)}\Bigl(l(s_A)+l(s_B)\Bigr).
\end{equation}
Since strategy \eqref{e:benevolent} is fully revealing (it is strictly increasing, as shown below), and under quadratic loss it is optimal for the voter to elect $A$ over $B$ if and only if
\[
(y_A(s_A)-y_B(s_B))\Bigl(\E[\theta \mid s_A,s_B]-\frac{y_A(s_A)+y_B(s_B)}{2}\Bigr) \geq 0,
\]
\autoref{e:benevolent-br-1} implies that it is optimal for the voter to elect $A$ if and only if
\begin{equation}
\label{e:benevolent-br-2}
(y(s_A)-y(s_B))\bigl(l(s_A)+l(s_B)\bigr)\ge 0.  
\end{equation}

Note that both $y\left(s_i\right)$ and $l\left(s_i\right)$ are odd and strictly increasing in $s_i$. Oddness follows from symmetry of the joint signal distribution around zero. For monotonicity, note that for $s_i>0$ the conditioning event $\abs{s_{-i}}\leq s_i$ can be written as $0\leq s_i+s_{-i}\leq 2s_i$, and equally as $0\leq s_i-s_{-i}\leq 2s_i$, so that
\[
y(s_i)=\frac{\beta}{\alpha+2\beta}\,\E\bigl[s_i+s_{-i} \mid s_i,\ 0\leq s_i+s_{-i}\leq 2s_i\bigr] \quad \text{and} \quad l(s_i)=\E\bigl[s_i-s_{-i} \mid s_i,\ 0\leq s_i-s_{-i}\leq 2s_i\bigr].
\]
Conditional on $s_i$, each of $s_i\pm s_{-i}$ is normally distributed with mean $\bigl(1\pm\frac{\beta}{\alpha+\beta}\bigr)s_i$, a positive multiple of $s_i$ in each case. Raising $s_i$ therefore raises each of these means and raises the upper truncation point $2s_i$, while the lower truncation point remains $0$; each such change strictly raises the expectation of a truncated normal.  Both expectations are of variables confined to $[0,2s_i]$, hence positive, so oddness extends strict monotonicity to all of $\Reals$.

Hence, when the realized $s_A$ and $s_B$ have the same sign, the term $l(s_A)+l(s_B)$ shares that sign, so whether inequality \eqref{e:benevolent-br-2} holds is determined by the sign of $y\left(s_A\right)-y\left(s_B\right)$. When instead the realized signals have opposite signs, $y(s_A)-y(s_B)$ has the same sign as $s_A$ (since $y$ is sign-preserving), so whether the inequality holds is determined by the sign of $l\left(s_A\right)+l\left(s_B\right)=l\left(s_A\right)-l\left(-s_B\right)$. By oddness and monotonicity of both $y$ and $l$, it follows that in both cases, inequality \eqref{e:benevolent-br-2} holds if and only if $\abs{s_A} \geq\abs{s_B}$, which---because $\abs{y(s)}$ is strictly increasing in $\abs{s}$---is equivalent to $\abs{y(s_A)}\geq \abs{y(s_B)}$. That is, the voter elects the candidate with the more extreme platform.
\end{proof}

\autoref{lem:benevolent-br} implies that if both candidates pander using strategy \eqref{e:benevolent} and the voter best responds, then the candidate with the more extreme signal wins. With that in mind, we now state the following result.

\begin{proposition}
\label{p:benevolent-best} Assume the normal-quadratic specification. Consider the symmetric strategy profile in which each
candidate $i$ panders by playing 
\eqref{e:benevolent} and the voter best responds. This profile maximizes voter welfare among all strategy profiles in which the voter's
best response would lead to candidate $i$ winning whenever $%
\abs{s_{i}}>\abs{s_{-i}}$. 
\end{proposition}

The intuition for \autoref{p:benevolent-best} %
is as follows. The welfare-maximizing platform given any information $%
\mathcal{I}$ is $\mathbb{E}[\theta  \mid \mathcal{I}]$. When the voter is
selecting the candidate with the most extreme signal, the relevant
information that candidate $i$ has when he conditions on winning is his own
signal, $s_{i}$, and that $\abs{s_{i}}>\abs{s_{-i}}$. %
Since the
voter would optimally elect the candidate with the most extreme signal if
both candidates used unbiased strategies, 
an implication of \autoref{p:benevolent-best} is that both candidates
playing the pandering strategy \eqref{e:benevolent} provides higher voter
welfare than both candidates playing unbiased strategies (and the voter best responding in each case), and hence also over any equilibrium---which is the statement of \autoref{p:benevolent-best-weaker}.

\begin{proof}[\textbf{Proof of \autoref{p:benevolent-best}}]
\hypertarget{proof:benevolent-best}{B}y the law of iterated expectations,
the voter's ex-ante utility can be expressed as 
\begin{align}
v(y_{A},y_{B},w_{A})& =-\mathbb{E}[(x-\theta )^{2}]=-\mathbb{E}[\mathbb{E}%
[\left( x-\theta \right) ^{2} \mid s_{A},s_{B}]]=-\mathbb{E}\left[ \left( x-\frac{%
\beta \left( s_{A}+s_{B}\right) }{\alpha +2\beta }\right) ^{2}\right] -\frac{%
1}{\alpha +2\beta }  \notag \\
& =-\Pr \left( A\ \text{wins}\right) \mathbb{E}\left[ \left( x_{A}-\frac{%
\beta \left( s_{A}+s_{B}\right) }{\alpha +2\beta }\right) ^{2}\Big |\ A\ 
\text{wins}\right]  \notag \\
& \qquad -\Pr \left( B\ \text{wins}\right) \mathbb{E}\left[ \left( x_{B}-%
\frac{\beta \left( s_{A}+s_{B}\right) }{\alpha +2\beta }\right) ^{2}\Big |\
B\ \text{wins}\right] -\frac{1}{\alpha +2\beta }.  \label{eqn:EU-1}
\end{align}
It is convenient to define $h_{i}(s_{i}) := \mathbb{E}\left[ s_{-i} \mid s_{i},i\ 
\text{wins}\right] $. Using iterated expectations again and a mean-variance
decomposition as in the \hyperlink{proof:uno}{proof of \autoref{uno}}, it
also holds that for any $i$, 
\begin{align}
& \mathbb{E}\left[ \left( x_{i}-\frac{\beta \left( s_{A}+s_{B}\right) }{%
\alpha +2\beta }\right) ^{2}\Big |\ i\ \text{wins}\right]  \notag \\
& \qquad =\mathbb{E}\left[ \mathbb{E}\left[ \left( x_{i}-\frac{\beta \left(
s_{A}+s_{B}\right) }{\alpha +2\beta }\right) ^{2}\Big |\ s_{i},i\ \text{wins}%
\right] \Big |\ i\ \text{wins}\right]  \notag \\
& \qquad =\mathbb{E}\left[ \left( x_{i}-\frac{\beta \left( s_{i}+\mathbb{E}%
\left[ s_{-i} \mid s_{i},i\ \text{wins}\right] \right) }{\alpha +2\beta }\right)
^{2}+\left( \frac{\beta }{\alpha +2\beta }\right) ^{2}\Var\left[
s_{-i} \mid s_{i},i\ \text{wins}\right] \Big |\ i\ \text{wins}\right]  \notag \\
& \qquad =\mathbb{E}\left[ \left( x_{i}-\frac{\beta \left(
s_{i}+h_i(s_{i})\right) }{\alpha +2\beta }\right) ^{2}\Big |\ i\ \text{wins}%
\right] +\left( \frac{\beta }{\alpha +2\beta }\right) ^{2}\mathbb{E}\left[ %
\Var\left[ s_{-i} \mid s_{i},i\ \text{wins}\right] \Big |\ i\ \text{wins}\right] .
\label{eqn:EU-2}
\end{align}

Equations \eqref{eqn:EU-1} and \eqref{eqn:EU-2} imply 
\begin{equation}
v(y_{A},y_{B},w_{A})=-\left( \frac{\beta }{\alpha +2\beta }\right)
^{2}L_{V}-L_{E}-\frac{1}{\alpha +2\beta }\text{,}  \label{decomp_all}
\end{equation}%
where 
\begin{align}
L_{V}& := \sum_{i=A,B}\Pr \left( i\ \text{wins}\right) \mathbb{E}\left[ \Var%
\left[ s_{-i} \mid s_{i},i\ \text{wins}\right] \Big |\ i\ \text{wins}\right] ,
\label{decomp_Lv} \\
L_{E}& := \sum_{i=A,B}\Pr \left( i\ \text{wins}\right) \mathbb{E}\left[
\left( y_{i}(s_{i})-\frac{\beta \left( s_{i}+h_i(s_{i})\right) }{\alpha
+2\beta }\right) ^{2}\Big |\ i\ \text{wins}\right] .  \label{decomp_LE}
\end{align}

Our problem is to maximize \eqref{decomp_all} subject to $i$ winning when $%
\abs{s_{i}}>\abs{s_{-i}}$. Since \eqref{decomp_Lv} does not depend on platforms
while \eqref{decomp_LE} is bounded below by $0$, a solution must satisfy for
each $i$: 
\begin{equation*}
y_{i}(s_{i})=\frac{\beta \left( s_{i}+h_i(s_{i})\right) }{\alpha +2\beta }=%
\mathbb{E}[\theta \mid s_{i},i\ \text{wins}].
\end{equation*}%
Since the constraint is that $i$ \text{wins} when $\abs{s_{i}}>\abs{s_{-i}}$, it
follows immediately that the solution is for each candidate to use the
strategy \eqref{e:benevolent}.
\end{proof}

We remark that although we do not have a proof, we conjecture that \autoref%
{p:benevolent-best} holds without the qualification that a candidate must
win when he has the more extreme signal.\footnote{\label{believe}For a
suggestive heuristic, consider any symmetric strategy profile in which both
candidates play the same strategy $y$ that is symmetric around $0$. For the
unbiased strategy, we have the derivative $y^{\prime }(\cdot )=\frac{\beta }{%
\beta +\alpha }$; for the overreaction strategy identified in \autoref%
{thm:FRE}, we have $y^{\prime }(\cdot )=\frac{2\beta }{\alpha +2\beta }$.
Presuming differentiability, one can verify that whenever $y^{\prime }(\cdot
)\in [0,\frac{2\beta }{\alpha +2\beta }]$, it would be optimal for
the voter to elect the candidate with the most extreme platform and hence
the most extreme signal. 
Thus, roughly speaking, the requirement that a candidate wins when he has
the most extreme signal is satisfied as long as neither candidate overreacts
by more than he would when conditioning on the opponent having received the
same signal as he did. It appears unlikely that such a degree of
overreaction could improve voter welfare.}

\subsection{Anti-Pandering Beyond the Normal-Normal Structure}

\label{app:exponential}

The existence of an anti-pandering equilibrium like the one characterized in \autoref{thm:FRE} holds beyond our
normal-normal informational structure. The simplest extension is to an asymmetric normal-normal specification in which, conditional on the state $\theta$, each candidate $i$ receives an independent signal $s_{i} \sim  \mathcal{N}(\theta,1/\beta_i)$, with different precisions $\beta_i$. %
In this case, the fully-revealing equilibrium strategy takes the form $y_{i}(s_{i})=\frac{2\beta _{i}}{\alpha +\beta _{A}+\beta _{B}}s_{i}$; candidate $i$ overreacts when $\alpha +\beta _{i}>\beta _{-i}$, so both candidates anti-pander when $\abs{\beta _{A}-\beta _{B}}<\alpha $.

More generally, maintaining quadratic loss voter preferences, a
fully-revealing anti-pandering equilibrium exists when the distributions of
the state $\theta $ and signals $s_{i}$ are conjugate and belong to a natural 
exponential family. The \hyperref[app:beta-bernoulli]%
{Supplementary Appendix}
explicitly derives such an equilibrium in a Beta-prior--Bernoulli-signals
specification and shows that it has characteristics analogous 
to that of \autoref{sec:pandering}. The key general property of a natural 
exponential family is that the posterior expectation $\mathbb{E}[\theta
 \mid s_{0},s_{1},\ldots ,s_{n}]$ of the state $\theta $ given a prior mean
parameter, say $s_{0}$, and any number of signal realizations, $s_{1},\ldots
,s_{n}$, is linear in $s_{0},s_{1},\ldots $ and $s_{n},$ \citep{Jewel:74}.
In our Downsian framework, suppose the two candidates' signals $s_{A}$ and $%
s_{B}$ are identically distributed conditional on the state $\theta $.
(Identical distributions are not necessary, but make the points below more
transparent.) Then, there are constants $w_{0}$ and $w_{1}$ such that 
\begin{equation*}
\mathbb{E}[\theta  \mid s_{i}]=\frac{w_{0}s_{0}+w_{1}s_{i}}{w_{0}+w_{1}}\text{ \
and \ }\mathbb{E}[\theta  \mid s_{A},s_{B}]=\frac{w_{0}s_{0}+2w_{1}\left( \left(
s_{A}+s_{B}\right) /2\right) }{w_{0}+2w_{1}}.
\end{equation*}
As a result, the following generalization of the existence result of \autoref%
{thm:FRE} can be verified:\footnote{As
the prior density need no longer be symmetric around the mean (unlike with a
normal prior) and signals may be bounded (unlike with normally distributed
signals), the definitions of anti-pandering or overreaction %
have to be broadened from earlier. We now say that a strategy 
$y_{i}$ has overreaction if for all $s_{i}$, $\abs{y_{i}(s_{i})-\mathbb{E}%
[\theta ]}\geq \abs{\mathbb{E}[\theta  \mid s_{i}]-\mathbb{E}[\theta ]}$ with strict
inequality for some $s_{i}$. %
The focus on posterior expectations of the state is justified when the voter
has a quadratic loss function. See the discussion in \citet{RouxSobel:12} to
get a sense of how asymmetric loss functions would affect the conclusions.}
there is an equilibrium with overreaction in which each candidate $i$ plays 
\begin{equation*}
y_{i}(s_{i})=\frac{2w_{1}}{w_{0}+2w_{1}}s_{i}+\frac{w_{0}}{w_{0}+2w_{1}}%
s_{0},
\end{equation*}%
and the voter randomizes uniformly after any pair of on-path platforms.%
\footnote{%
While there may now be off-path platforms (unlike with normal
distributions), as in the Beta-Bernoulli example in the \hyperref[app:beta-bernoulli]%
{Supplementary Appendix}, the equilibrium can be supported with reasonable
off-path beliefs.} 

\subsection{The Welfare Bound with Mixed Strategies}
\label{app:mixing}

Here we extend \autoref{thm:welfare} to equilibria in which candidates randomize. Each candidate $i$ now uses a measurable mixed strategy $\sigma _{i}:S_{i}\rightarrow \Delta (X)$ satisfying $\E[\abs{u(x_{i},\theta )}]<\infty$.

\begin{proposition}
\label{p:mixing}
Assume the signal distribution $F_{s_{A},s_{B}}$ satisfies \autoref{completeness_general}, and consider any equilibrium with candidates' strategies $(\sigma _{A},\sigma _{B})$. There is a candidate $i\in \{A,B\}$ such that the voter's welfare in this equilibrium is the same as if candidate $i$ were elected no matter which policies are proposed using $(\sigma_A,\sigma_B)$.
\end{proposition}

The proposition does not assume either strategy is identifiable, so \autoref{thm:expost-general} need not apply and the equilibrium need not be ex post, as discussed in the first part of \autoref{sec:elections-discussion}.

\begin{proof}
Fix an equilibrium. Let $\mu _{i}$ denote the ex-ante distribution of candidate $i$'s platform, $U^{\ast }_{i}$ denote $i$'s probability of winning, $v$ denote the voter's welfare, and $v_{i}:=\E[u(x_{i},\theta )]$ denote the voter's welfare from electing $i$ no matter the platforms. Define candidate $i$'s win probability with platform $x_{i}$ when his opponent's signal is $s_{-i}$ as
\begin{equation*}
\phi _{i}(x_{i},s_{-i}):=\int w_{i}(x_{i},x_{-i})\,\sigma _{-i}(\mathrm{d}x_{-i}\mid s_{-i}).
\end{equation*}

By \autoref{lem:zerosum_general} and iterated expectations, $\E_{s_{-i}}[\phi _{i}(x_{i},s_{-i})-U^{\ast }_{i}\mid s_{i}]=0$ for $\mu _{i}$-a.e.~$x_{i}$ and $F_{s_{i}}$-a.e.~$s_{i}$. \autoref{completeness_general} then implies $\phi _{i}(x_{i},s_{-i})=U^{\ast }_{i}$ for $(\mu _{i}\otimes F_{s_{-i}})$-a.e.~$(x_{i},s_{-i})$. Given the maintained assumption that $F_{s_{A},s_{B}}$ is absolutely continuous with respect to $F_{s_{A}}\otimes F_{s_{B}}$, the joint distribution of $i$'s platform and his opponent's signal is absolutely continuous with respect to $\mu _{i}\otimes F_{s_{-i}}$, by the argument of \autoref{fn:domination} with $s_{-i}$ in place of $x_{-i}$. Hence $\phi _{i}(x_{i},s_{-i})=U^{\ast }_{i}$ a.s. This implies
\begin{equation*}
\E\left[ w_{i}(x_{i},x_{-i})\,u(x_{i},\theta )\right] =\E\left[ \phi _{i}(x_{i},s_{-i})\,u(x_{i},\theta )\right] =U^{\ast }_{i}v_{i},
\end{equation*}
where the first equality is from averaging over $x_{-i}$, which given $s_{-i}$ is independent of $x_{i}$ and $\theta$. 

Summing over both candidates gives $v=U^{\ast }_{A}v_{A}+U^{\ast }_{B}v_{B}\leq \max\{v_A,v_B\}$, where the inequality is because $U^{\ast }_{A}+U^{\ast }_{B}=1$. But we also have $v\geq \max \{v_{A},v_{B}\}$, because the voter can always choose to elect a single candidate regardless of the platforms. Thus $v=\max \{v_{A},v_{B}\}$.
\end{proof}

\section{Completeness and Strong Linear Independence}
\label{app:SLI}

Using ``signal'' as a synonym for ``type'', recall that we noted in the main text after introducing \autoref{completeness_general} that it is equivalent to a full row and column rank condition for finite signal spaces. This appendix clarifies that relationship more generally, with its main result being \autoref{prop:comp_sli_equiv} below. The setting and notation for this appendix follow \autoref{sec:zerosum}. 

\subsection{Strong Linear Independence}

Let $\mathcal B(S_i)$ denote the Borel $\sigma$-algebra on $S_i$ and let $\mathcal M(S_i)$ denote the set of finite signed measures on $S_i$.
Define the linear operator $K_i:\mathcal M(S_i)\to \mathcal M(S_{-i})$ by setting, for any $\mu\in \mathcal M(S_i)$ and $B \in \mathcal{B}(S_{-i})$,
\[
(K_i \mu)(B)
\;:=\;
\int_{S_i} F(B \mid s_i)\,\mu(\mathrm ds_i).
\]
For any $\mu\in \mathcal M(S_i)$, the map $s_i \mapsto F(B\mid s_i)$ is bounded and measurable, so $(K_i\mu)(B)$ is a
well-defined Lebesgue integral for each $B \in \mathcal{B}(S_{-i})$. Moreover, $K_i\mu$ is a finite signed measure because $\mu$ is finite and $F(\cdot\mid s_i)$ is a probability measure for each $s_i$.

Let $\mathcal{M}_{ac}(S_i)\subset \mathcal{M}(S_i)$ denote the subset of finite signed measures that are dominated by (i.e., absolutely continuous with respect to) the marginal $F_i$. Writing, as usual, that a measure $\mu=0$ means $\mu(\cdot)=0$, we define:

\begin{definition}
\label{def:SLI}
The family $\{F(\cdot \mid s_i)\}_{s_i \in S_i}$ is \emph{strongly linearly independent (SLI) relative to $F_i$} if for every $\mu \in \mathcal M_{ac}(S_i)$,
\[
K_i \mu = 0 \implies  \mu = 0.
\]
\end{definition}

For brevity, we will sometimes refer to the condition as just SLI, leaving the marginal $F_i$ implicit. The marginal affects SLI only through its null sets: replacing $F_i$ with any measure having the same null sets leaves $\mathcal{M}_{ac}(S_i)$, and hence the definition, unchanged. The null-sets qualification is unavoidable because the conditional distributions are themselves pinned down only up to $F_i$-null sets; it also matches the a.e.~nature of completeness defined below. Requiring $K_i\mu=0\implies\mu=0$ for all finite signed measures would make the property strictly stronger in general, and it would break the equivalence of \autoref{prop:comp_sli_equiv} below.

SLI requires that no nonzero finite signed measure that is dominated by $F_i$ creates a zero mixture of the conditional distributions $\{F(\cdot\mid s_i)\}_{s_i\in S_i}$. When $S_i$ is finite, SLI is equivalent to linear independence. When $S_i$ is countable and $F_i$ assigns positive probability to every signal, every finite signed measure is dominated by $F_i$, so SLI is stronger than textbook linear independence because SLI does not restrict to finite mixtures.\footnote{Recall that even when $S_i$ is infinite, $\{F(\cdot \mid s_i)\}_{s_i \in S_i}$ is \emph{linearly independent} if every nonzero \emph{finite} linear combination is nontrivial, i.e., for any finite set
 $\{s_i^1,\ldots,s_i^K\}\subset S_i$ and any $c:\{1,\ldots,K\}\to \Reals$, it holds that
$$
\sum_{k=1}^K c(k) F(\cdot \mid s_i^k) =0 \implies c(\cdot)=0.
$$
} 
Allowing for infinite mixtures is needed for the equivalence with completeness (\autoref{prop:comp_sli_equiv} below), as demonstrated by \autoref{eg:LI-notSLI}---there, every nontrivial finite signed combination of the conditional distributions is nonzero, yet an infinite one vanishes.

\subsection{Completeness}

As usual, write $L^1(F_{i})$ for the measurable and $F_{i}$-integrable real-valued functions on $S_{i}$.

\begin{definition}
\label{def:completeness_appendix}
Player $i$'s signals have a \emph{complete} family of conditional distributions \mbox{$\{F(\cdot\mid s_i)\}_{s_i\in S_i}$} if for every $g\in L^1(F_{-i})$ it holds that
\[
\mathbb{E}[g(s_{-i}) \mid s_{i}] = 0
\text{ for } F_i\text{-a.e.~} s_i
\implies
F\left(g(s_{-i})=0 \mid s_i\right) = 1 \text{ for } F_i\text{-a.e.~} s_i.
\]
\end{definition}

For short, we will say that there is \emph{completeness for $i$} when player $i$'s signals have a complete family of conditional distributions. Completeness (when required for both players) is closely related to \autoref{completeness_general} and captures the same idea; it is, however, slightly stronger in general because it allows for unbounded test functions $g$. Unbounded test functions are irrelevant when $S_{-i}$ is finite, but we will see that they are essential for the connection of completeness to SLI when $S_{-i}$ is infinite. When \autoref{def:completeness_appendix} is restricted to bounded test functions $g$, we say that there is \emph{bounded completeness} for $i$.

\subsection{Equivalence}

Completeness for $i$ concerns how informative $s_i$ is about the \emph{other player} $-i$'s signal. So, even with finite signal sets, completeness for $i$ corresponds to linear independence of $\{F(\cdot \mid s_{-i})\}_{s_{-i}\in S_{-i}}$, not of $i$'s own conditional distributions. Indeed, with finite signal sets for both players (albeit of different cardinality), we can have completeness for $i$ but a failure of linear independence of $\{F(\cdot \mid s_{i})\}_{s_{i}\in S_{i}}$, and vice-versa.\footnote{In particular, linear independence fails for $i$ when $S_i$ has duplicate signals, while there is completeness for $i$ when $\abs{S_{-i}}=1$. Conversely, if $\abs{S_{-i}}>\abs{S_i}=1$, there is linear independence for $i$ but not completeness for $i$.}

With that in mind, we have the following equivalence.

\begin{proposition}
\label{prop:comp_sli_equiv}
For each player $i\in\{A,B\}$, it holds that
\[
\text{Completeness for $i$}
\iff
\text{SLI relative to $F_{-i}$}.
\]
\end{proposition}

\begin{proof}
Fix an arbitrary player $i\in\{A,B\}$ and write $j$ for $-i$.

\medskip
\noindent\textbf{(Completeness for $i$ $\boldsymbol{\implies}$ SLI relative to $F_j$).}
Assume completeness for $i$. Let $\mu\in \mathcal M_{ac}(S_j)$ such that $K_j\mu = 0$. We must show that $\mu=0$.

Let $g(s_j) := \mathrm{d}\mu/\mathrm d F_j(s_j)$ be the Radon--Nikodym derivative of $\mu$ with respect to $F_j$, which exists because $\mu$ is dominated by $F_j$.
Fix any measurable $A \subset S_i$. Since \mbox{$F(A\mid s_j)=\mathbb{E}[\ind_A(s_i)\mid s_j]$}, the law of iterated expectations gives
\begin{align*}
(K_{j}\mu)(A)
&= \int_{S_{j}} F(A \mid s_{j})\,g(s_{j})\,F_{j}(\mathrm ds_{j})
= \mathbb{E}\!\left[\,\ind_{A}(s_{i})\,g(s_{j})\,\right]= \int_{A}\mathbb{E} [\,g(s_{j}) \mid s_{i}\,]\,F_{i}(\mathrm ds_{i}).
\end{align*}
Since $K_j\mu = 0$, the last integral vanishes for all measurable $A\subset S_i$, and hence \mbox{$\mathbb{E}[\,g(s_j)\mid s_i\,]=0$} for $F_i$-a.e.~$s_i$. By completeness for $i$, it follows that $F(g(s_j)=0 \mid s_i)=1$ for $F_i$-a.e.~$s_i$.
By the law of total probability, $g(s_j)=0$ $F_j$-a.s., and therefore $\mu=0$.

\medskip
\noindent\textbf{(SLI relative to $F_j$ $\boldsymbol{\implies}$ Completeness for $i$).}
Assume SLI relative to $F_j$. Let $g\in L^1(F_j)$ satisfy
\begin{equation}
\label{e:SLI-Complete-1}
\mathbb{E}[\,g(s_j) \mid s_i\,] = 0 \text{ for }F_i\text{-a.e.~} s_i.
\end{equation}
Define $\mu\in \mathcal{M}_{ac}(S_j)$ by
\begin{equation}
\label{e:SLI-Complete-2}
\mu(B) := \int_B g(s_j)\,F_j(\mathrm{d}s_j) \quad \text{ for any $B \in \mathcal{B}(S_{j})$}.    
\end{equation}
Fix any measurable $A\subset S_i$. Since $F(A\mid s_j)=\mathbb{E}[\ind_A(s_i)\mid s_j]$, the law of iterated expectations gives
\begin{align*}
(K_j \mu)(A)
&= \int_{S_j} F(A \mid s_j)\, g(s_j)\,F_j(\mathrm{d}s_j)
= \mathbb{E}\!\left[\,\ind_A(s_i)\,g(s_j)\,\right]
= \int_{A}\mathbb{E} [\,g(s_{j}) \mid s_{i}\,]\,F_{i}(\mathrm ds_{i}) = 0,
\end{align*}
where the last equality is by \eqref{e:SLI-Complete-1}.
Thus $K_j\mu=0$. SLI implies $\mu=0$, which by \eqref{e:SLI-Complete-2} implies $g(s_j)=0$ $F_j$-a.s. %
The law of total probability now implies $F\!\left(g(s_j)=0 \mid s_i\right)=1$ for $F_i$-a.e.~$s_i$, which is completeness for $i$.
\end{proof}

\autoref{prop:comp_sli_equiv} formalizes the sense in which SLI is the appropriate infinite-dimensional analog of full rank (of the other player's conditional distributions) for completeness. %
The following countable-signal example shows both why the relevant directions of \autoref{prop:comp_sli_equiv} use completeness rather than bounded completeness, and why finite-mixture linear independence is weaker than SLI. 

\begin{example}
\label{eg:LI-notSLI}
Consider $S_A=\{0,1, \ldots\}$ and $S_B=\{1,2, \ldots\}$. While we could use any signal distribution $F$ that has a full-support marginal $F_A$, for concreteness take $F_A(0)=1/2$ and $F_A(s_A)=3^{-s_A}$ for $s_A>0$, and let the conditional distributions be
$$
F(s_B \mid s_A)=\begin{cases}
    2^{-s_B} & \text{ if } s_A=0\\
    \ind\{s_B=s_A\} & \text{ if } s_A>0.
\end{cases}
$$

Bounded completeness holds for player $B$: for any signal $s_B$, the conditional distribution $F(\cdot  \mid  s_B)$ is supported on $\{0, s_B\}$; thus, if a bounded function $g(s_A)$ has $\E[g(s_A) \mid s_B]=0$ for all $s_B$, then $g(0)=0$ by considering large enough $s_B$, and consequently $g(\cdot)=0$.%

The family $\{F(\cdot  \mid  s_A)\}_{s_A\in S_A}$ does not satisfy SLI because for any $s_B\in S_B$, we have
$$F(s_B \mid 0)+\sum_{s_A=1}^\infty \left(-2^{-s_A}\right) F(s_B\mid s_A)=2^{-s_B}-2^{-s_B}=0.$$ The family $\{F(\cdot  \mid  s_A)\}_{s_A\in S_A}$ is, however, linearly independent: if $\sum_{s_A=0}^K c(s_A) F(\cdot\mid s_A)=0$ for any integer $K$, then $c(0)=0$ by considering $s_B>K$, and consequently $c(\cdot)=0$. 

These observations show that bounded completeness for $B$ does not imply SLI relative to $F_A$, and, using \autoref{prop:comp_sli_equiv}, linear independence for $A$ does not imply completeness for $B$.\footnote{The failure of completeness can also be directly verified using %
the test function $g$ given by $g\left(0\right)=1$ and $g(s_A)=-(1/2)(3/2)^{s_A}$ for $s_A>0$.
}
\end{example}

\subsection{Insufficiency of Convex Independence}

In their work on informational richness in mechanism design, \citet{CM88} discussed the role of both linear independence (for dominant-strategy mechanisms) and also {convex independence} (for Bayesian-incentive-compatible mechanisms) with finite types; \citet{MR92} extended the latter analysis to an infinite setting.  

Even with finite signals, convex independence---being weaker than linear independence---is insufficient for our purposes. Formally, say that \emph{convex independence} holds for player $i\in \{A,B\}$ if for all $s_i\in S_i$, $$F(\cdot \mid s_i) \notin \co\left(\{F(\cdot \mid t_i):t_i \in S_i\setminus \{s_i\}\}\right),$$ where $\co(\cdot)$ denotes the convex hull. 

Consider the following joint distribution when each player has 4 signals:
\[
F
=
\frac{1}{40}
\begin{pmatrix}
4 & 2 & 1 & 3\\
2 & 4 & 3 & 1\\
1 & 3 & 4 & 2\\
3 & 1 & 2 & 4
\end{pmatrix}.
\]
This is a symmetric distribution in which the marginal distribution of a player's signal is uniform. Hence, the matrix of conditional distributions for either player is just a rescaling of the above matrix (multiplying it by $4$). For either player $i$, the family of conditional distributions $\{F(\cdot \mid s_i)\}_{s_i\in S_i}$ is convexly independent---each conditional distribution $F(\cdot \mid s_i)$ assigns highest probability to a distinct opponent signal---yet  completeness (or equivalently here, bounded completeness) fails because the conditional matrix has rank $3$ (as the sum of the first and third rows of the $F$ matrix equals the sum of the second and fourth rows).

The conclusion of \autoref{thm:expost-general} fails under this joint distribution:

\begin{example}
\label{eg:CI}
Consider the above signal structure, labeling signals as $S_i=\{1,2,3,4\}$ in the natural way. Let $X_A=X_B=\{0,1\}$ and $
u_A(x_A,x_B)=-\ind\{x_A=x_B\}$. There is an identifiable equilibrium in which each player takes action $1$ when he receives signal $1$ or $3$, and takes action $0$ otherwise; the signal structure implies that each type of each player then faces a uniform distribution over the opponent's actions. This is, however, not an ex-post equilibrium.
\end{example}

\section{On \autoref{cor:singleoutcome}}
\label{app:Kattwinkel}
Here we make precise the connection between   \autoref{cor:singleoutcome} and \citet[Proposition 3, part 2]{Kattwinkel22}. They consider direct mechanisms, so $X_i=S_i$, and an outcome space $\Omega=[0,1]$, interpreted as an allocation probability. Their primitive is preferences over outcomes given by $\tilde u_A(\omega)=\omega$ and $\tilde u_B(\omega)=-\omega$. Given any mechanism (or outcome function) $w$, the induced preferences are  $u_i(x_A,x_B):=\tilde u_i(w(x_A,x_B))$, and hence there are strict preferences over outcomes. \citeauthor{Kattwinkel22} ask which mechanisms are incentive compatible in the sense that truthful reporting (i.e., the strategy $s_i\mapsto s_i$) forms an equilibrium. Since such strategies are identifiable (being pure), \autoref{cor:singleoutcome} implies that under \autoref{completeness_general} only constant mechanisms are incentive compatible. This subsumes \citet[Proposition 3, part 2]{Kattwinkel22}, which assumes finite type sets, in which case \autoref{completeness_general} reduces to their full-rank condition.\footnote{To be more precise: \autoref{completeness_general} is equivalent to full rank when $\abs{S_A}=\abs{S_B}<\infty$. Since \autoref{completeness_general} is stated as applying to both players, it cannot hold with finite type sets when $\abs{S_A}\neq \abs{S_B}$, because it requires full row and column rank. However, as seen in the proof of \autoref{thm:expost-general}, the theorem---and hence also \autoref{cor:singleoutcome}---only requires that \autoref{completeness_general} hold for one player $i$ when player $-i$'s equilibrium strategy is identifiable. So when both players' strategies are identifiable, as in the present discussion, it is sufficient that \autoref{completeness_general} hold for either player. With finite type sets, that reduces to usual full rank (i.e., {either} full row or column rank).} 

In the other direction, \citepos{Kattwinkel22} result can be combined with a revelation-principle argument to derive Corollary 1 for finite type sets and pure-strategy equilibria. However, to deal with infinite type sets or (identifiable) mixed-strategy equilibria, we believe an argument like ours is needed.
    
\setstretch{1.25}
\bibliographystyle{aer}
\bibliography{KST}

@article{MOS:20,
	Author = {Antony Millner and H\'elène Ollivier and Leo Simon},
	Journal = {Journal of Public Economics},
	Pages = {104175},
	Title = {Confirmation Bias and Signaling in Downsian Elections},
	Volume = {185},
	Year = {2020}}

@article{KvW:17,
	Author = {Kartik, Navin and Van Weelden, Richard},
	Journal = {Review of Economic Studies},
	Pages = {755--784},
	Title = {Informative Cheap Talk in Elections},
	Volume = {86},
    Number = {2},
	Year = {2019}}

@article{ABK:16,
	Author = {Ambrus, Attila and Volodymyr Baranovskyi and Aaron Kolb},
	Journal = {American Economic Journal: Microeconomics},
	Number = {4},
	Pages = {373--419},
	Title = {A Delegation-Based Theory of Expertise},
	Volume = {13},
	Year = {2021}}

@incollection{PC:09,
	Author = {P{\'e}try, Fran{\c{c}}ois and Collette, Beno{\^\i}t},
	Booktitle = {Do They Walk Like They Talk? Speech and Action in Policy Processes},
	Chapter = {5},
	Editor = {Louis M. Imbeau},
	Pages = {65--80},
	Publisher = {Springer},
	Title = {Measuring how political parties keep their promises: a positive perspective from political science},
	Year = {2009}}

@article{Alesina:88,
	Author = {Alesina, Alberto},
	Journal = {American Economic Review},
	Month = {September},
	Number = {4},
	Pages = {796--805},
	Title = {Credibility and Policy Convergence in a Two-Party System with Rational Voters},
	Volume = {78},
	Year = {1988}}

@article{CW:07,
	Author = {Steven Callander and Simon Wilkie},
	Journal = {Games and Economic Behavior},
	Number = {2},
	Pages = {262--286},
	Title = {Lies, Damned Lies, and Political Campaigns},
	Volume = {60},
	Year = {2007}}

@article{Campante:11,
	Author = {Filipe R. Campante},
	Journal = {Journal of Public Economics},
	Number = {7},
	Pages = {646--656},
	Title = {Redistribution in a model of voting and campaign contributions},
	Volume = {95},
	Year = {2011}}

@article{AES:13,
	Author = {Daron Acemoglu and Georgy Egorov and Konstantin Sonin},
	Journal = {Quarterly Journal of Economics},
	Number = {2},
	Pages = {771--805},
	Title = {A Political Theory of Populism},
	Volume = {128},
	Year = {2013}}

@article{Hotelling:29,
	Author = {Harold Hotelling},
	Journal = {Economic Journal},
	Month = {March},
	Number = {153},
	Pages = {41--57},
	Title = {Stability in Competition},
	Volume = {39},
	Year = {1929}}

@article{Gratton:14,
	Author = {Gabriele Gratton},
	Journal = {Games and Economic Behavior},
	Pages = {163--179},
	Title = {Pandering and Electoral Competition},
	Volume = {84},
	Year = {2014}}

@article{AP:02,
	Author = {Enriqueta Aragones and Thomas Palfrey},
	Journal = {Journal of Economic Theory},
	Pages = {131--161},
	Title = {Mixed Equilibrium in a Downsian Model with a Favored Candidate},
	Volume = {103},
	Year = {2002}}

@book{Downs:57,
	Address = {New York},
	Author = {Downs, Anthony},
	Publisher = {Harper and Row},
	Title = {An Economic Theory of Democracy},
	Year = {1957}}

@article{BC:97,
	Author = {Tim Besley and Stephen Coate},
	Journal = {Quarterly Journal of Economics},
	Pages = {85--114},
	Title = {An Economic Model of Representative Democracy},
	Volume = {112},
	Year = {1997}}

@article{Groseclose:01,
	Author = {Timothy Groseclose},
	Journal = {American Journal of Political Science},
	Pages = {862--886},
	Title = {A Model of Candidate Location when One Candidate Has a Valence Advantage},
	Volume = {45},
	Year = {2001}}

@article{OS:96,
	Author = {Martin J. Osborne and Al Slivinski},
	Journal = {Quarterly Journal of Economics},
	Pages = {65--96},
	Title = {A Model of Political Competition with Citizen-Candidates},
	Volume = {111},
	Year = {1996}}

@article{Palfrey:84,
	Author = {Thomas Palfrey},
	Journal = {Review of Economic Studies},
	Pages = {139--156},
	Title = {Spatial Equilibrium with Entry},
	Volume = {51},
	Year = {1984}}

@article{Wittman:83,
	Author = {Donald Wittman},
	Journal = {American Political Science Review},
	Pages = {142--157},
	Title = {Candidate Motivation: A Synthesis of Alternatives},
	Volume = {77},
	Year = {1983}}

@unpublished{Klumpp:11,
	Author = {Tilman Klumpp},
	Month = {August},
	Note = {unpublished},
	Title = {Populism, Partisanship, and the Funding of Political Campaigns},
	Url = {http://www.ualberta.ca/~klumpp/docs/populism.pdf},
	Year = {2014}}

@book{Zaller:92,
	Address = {New York},
	Author = {John R. Zaller},
	Publisher = {Cambridge University Press},
	Title = {The Nature and Origins of Mass Opinion},
	Year = {1992}}

@article{Gilens:01,
	Author = {Martin Gilens},
	Journal = {American Political Science Review},
	Number = {2},
	Pages = {379--396},
	Title = {Political Ignorance and Collective Policy Preferences},
	Volume = {95},
	Year = {2001}}

@article{Althaus:98,
	Author = {Scott L. Althaus},
	Journal = {American Political Science Review},
	Number = {3},
	Pages = {545--558},
	Title = {Information Effects in Collective Preferences},
	Volume = {92},
	Year = {1998}}

@book{Fishkin:97,
	Address = {New Haven},
	Author = {James S. Fishkin},
	Publisher = {Yale University Press},
	Title = {The Voice of the People: Public Opinion and Democracy},
	Year = {1997}}

@book{IK:87,
	Address = {Chicago},
	Author = {Shanto Iyengar and Donald Kinder},
	Publisher = {University of Chicago Press},
	Title = {News That Matters: Television and American Opinion},
	Year = {1987}}

@book{CCMS:60,
	Address = {Chicago},
	Author = {Angus Campbell and Philip E. Converse and Warren E. Miller and Donald E. Stokes},
	Publisher = {University of Chicago Press},
	Title = {The American Voter},
	Year = {1960}}

@book{SP:81,
	Author = {Howard Schuman and Stanley Presser},
	Publisher = {Academic Press},
	Title = {Questions and Answers in Attitude Surveys: Experiments on Question Form, Wording, and Context},
    Address = {New York},
	Year = {1981}}

@article{CDK:11,
	Author = {Yeon-Koo Che and Wouter Dessein and Navin Kartik},
	Journal = {American Economic Review},
	Month = {February},
	Number = 1,
	Pages = {47--79},
	Title = {Pandering to Persuade},
	Volume = 103,
	Year = {2013}}

@article{Honryo:18,
	author = {Takakazu Honryo},
	journal = {Journal of Economic Theory},
	pages = {273--87},
	title = {Risky Shifts as Multi-sender Signaling},
	volume = {174},
	year = {2018},
    }

@article{HL:03,
	Author = {Paul Heidhues and Johan Lagerlof},
	Journal = {Games and Economic Behavior},
	Month = {January},
	Number = {1},
	Pages = {48--74},
	Title = {Hiding Information in Electoral Competition},
	Volume = {42},
	Year = {2003}}

@article{MT:04,
	Author = {Eric Maskin and Jean Tirole},
	Journal = {American Economic Review},
	Month = {September},
	Number = {4},
	Pages = {1034--1054},
	Title = {The Politician and the Judge: Accountability in Government},
	Volume = {94},
	Year = {2004}}

@article{OS:06,
	Author = {Marco Ottaviani and Peter Norman Sorensen},
	Journal = {RAND Journal of Economics},
	Month = {Spring},
	Number = {1},
	Pages = {155--175},
	Title = {Reputational Cheap Talk},
	Volume = {37},
	Year = {2006}}

@article{MM:04,
	Author = {Sumon Majumdar and Sharun W. Mukand},
	Journal = {American Economic Review},
	Month = {September},
	Number = {4},
	Pages = {1207--1222},
	Title = {Policy Gambles},
	Volume = {94},
	Year = {2004}}

@unpublished{Loertscher:10,
	Author = {Simon Loertscher},
	Note = {mimeo, University of Melbourne},
	Title = {Location Choice and Information Transmission},
	Year = {2012}}

@article{CWHS:01,
	Author = {Brandice Canes-Wrone and Michael Herron and Kenneth W. Shotts},
	Journal = {American Journal of Political Science},
	Month = {July},
	Number = {3},
	Pages = {532--550},
	Title = {Leadership and Pandering: A Theory of Executive Policymaking},
	Volume = {45},
	Year = {2001}}

@article{BDS:09,
	Author = {Bernhardt, Dan and Duggan, John and Squintani, Francesco},
	Journal = {Journal of Economic Theory},
	Month = {September},
	Number = {5},
	Pages = {2021--2056},
	Title = {Private Polling in Elections and Voter Welfare},
	Volume = {144},
	Year = 2009}

@article{MM:02,
	Author = {Martinelli, Cesar and Matsui, Akihiko},
	Journal = {Journal of Public Economic Theory},
	Number = {1},
	Pages = {39--61},
	Title = {Policy Reversals and Electoral Competition with Privately Informed Parties},
	Volume = {4},
	Year = 2002}

@article{Schultz:96,
	Author = {Schultz, Christian},
	Journal = {Review of Economic Studies},
	Month = {April},
	Number = {2},
	Pages = {331--44},
	Title = {Polarization and Inefficient Policies},
	Volume = {63},
	Year = 1996}

@article{LVdS:04,
	Author = {Jean-Fran\c{c}ois Laslier and Van de Straeten, Karine},
	Journal = {Economic Theory},
	Month = {August},
	Number = {2},
	Pages = {419-446},
	Title = {Electoral Competition under Imperfect Information},
	Volume = {24},
	Year = 2004}

@article{Martinelli:01,
	Author = {Martinelli, Cesar},
	Journal = {Public Choice},
	Month = {July},
	Number = {1--2},
	Pages = {147-67},
	Title = {Elections with Privately Informed Parties and Voters},
	Volume = {108},
	Year = 2001}

@article{BDS:07,
	Author = {Bernhardt, Dan and Duggan, John and Squintani, Francesco},
	Journal = {Games and Economic Behavior},
	Month = {January},
	Number = {1},
	Pages = {1--29},
	Title = {Electoral competition with privately-informed candidates},
	Volume = {58},
	Year = 2007}

@article{KM:07,
	Author = {Navin Kartik and R. Preston McAfee},
	Journal = {American Economic Review},
	Month = {June},
	Number = {3},
	Pages = {852--870},
	Title = {Signaling Character in Electoral Competition},
	Volume = {97},
	Year = {2007}}

@article{PS:96,
	Author = {Canice Prendergast and Lars Stole},
	Journal = {Journal of Political Economy},
	Month = {December},
	Number = {6},
	Pages = {1105--34},
	Title = {Impetuous Youngsters and Jaded Old-Timers: Acquiring a Reputation for Learning},
	Volume = {104},
	Year = 1996}

@article{Levy:04,
	Author = {Gilat Levy},
	Journal = {European Economic Review},
	Month = {June},
	Number = {3},
	Pages = {503--525},
	Title = {Anti-herding and Strategic Consultation},
	Volume = {48},
	Year = 2004}

@article{GS:09,
	Author = {Edward L. Glaeser and Cass R. Sunstein},
	number = {1},
	Journal = {Journal of Legal Analysis},
	Month = {Winter},
	Pages = {263--324},
	Title = {Extremism and Social Learning},
	Volume = {1},
	Year = {2009}}

@article{Calvert:85,
	Author = {Randall L. Calvert},
	Journal = {American Journal of Political Science},
	Pages = {69--95},
	Title = {Robustness of the Multidimensional Voting Model: Candidate Motivations, Uncertainty, and Convergence},
	Volume = {29},
	Year = {1985}}

@article{Jewel:74,
	Author = {William S. Jewell},
	Journal = {ASTIN Bulletin},
	Pages = {77--90},
	Title = {Credible Means are Exact Bayesian for Exponential Families},
	Volume = {8},
	Year = {1974}}

@article{Wittman:89,
	Author = {Donald Wittman},
	Journal = {Journal of Political Economy},
	Pages = {1395--1424},
	Title = {Why Democracies Produce Efficient Results},
	Volume = {97},
	Year = {1989}}

@article{RouxSobel:12,
	Author = {Nicolas Roux and Joel Sobel},
	Month = {November},
    journal = {American Economic Journal: Microeconomics},
	Title = {Group Polarization in a Model of Information Aggregation},
	Year = {2015},
    Volume = 7,
    number = 4,
    pages = {202--32}
    }

@article{Glaeser:05,
	Author = {Edward L. Glaeser and Giacomo A. M. Ponzetto and Jesse M. Shapiro},
	Journal = {Quarterly Journal of Economics},
	Month = {November},
	Number = {4},
	Pages = {1283--1330},
	Title = {Strategic Extremism: Why Republicans and Democrats Divide on Religious Values},
	Volume = {120},
	Year = 2005}

@techreport{Viossat:06,
	Author = {Viossat, Yannick},
	Institution = {Stockholm School of Economics},
	Month = Aug,
	Number = {641},
	Title = {{The Geometry of Nash Equilibria and Correlated Equilibria and a Generalization of Zero-Sum Games}},
	Type = {SSE/EFI Working Paper Series in Economics and Finance},
	Url = {http://ideas.repec.org/p/hhs/hastef/0641.html},
	Year = 2006}

@book{Lehmann:86,
  title={Testing Statistical Hypotheses},
  author={Lehmann, Erich L.},
  year={1986},
  publisher={Wiley},
  edition = {2nd}
}

@article{LPP:23,
    author = {{Le Pennec}, Caroline and Pons, Vincent},
    title = {How do Campaigns Shape Vote Choice? Multicountry Evidence from 62 Elections and 56 TV Debates},
    journal = {Quarterly Journal of Economics},
    volume = {138},
    number = {2},
    pages = {703--767},
    year = {2023},
}

@article{VN:28,
  title={On the Theory of Games of Strategy},
  author={{Von Neumann}, John},
  journal={Mathematische Annalen},
  volume={100},
  pages={295--320},
  year={1928}
}

@article{Nash:51,
 author = {John Nash},
 journal = {Annals of Mathematics},
 number = {2},
 pages = {286--295},
 title = {Non-Cooperative Games},
 volume = {54},
 year = {1951}
}

@article{JMMZ06,
  title     = {The Limits of Ex Post Implementation},
  author    = {Jehiel, Philippe and Meyer-ter-Vehn, Moritz and Moldovanu, Benny and Zame, William R.},
  journal   = {Econometrica},
  volume    = {74},
  number    = {3},
  pages     = {585--610},
  year      = {2006},
  doi       = {10.1111/j.1468-0262.2006.00675.x}
}

@article{PS24,
  title     = {Robust Implementation with Costly Information},
  author    = {Pei, Harry and Strulovici, Bruno},
  journal   = {Review of Economic Studies},
  year      = {2025},
  pages = {476--505},
  volume   = {92},
  number = {1},
  month = {January},
}

@article{Bils23,
    year = {2023},
    volume = {18},
    journal = {Quarterly Journal of Political Science},
    title = {Overreacting and Posturing: How Accountability and Ideology Shape Executive Policies},
    doi = {10.1561/100.00020177},
    number = {2},
    pages = {153--182},
    author = {Peter Bils}
}

@misc{Kattwinkel22,
  author = {Kattwinkel, Deniz and Niemeyer, Axel and Preusser, Justus and Winter, Alexander},
  title  = {Mechanisms without Transfers for Fully Biased Agents},
  year   = {2022},
  note   = {arXiv:2205.10910}
}

@unpublished{KST15,
  author = {Navin Kartik and Francesco Squintani and Katrin Tinn},
  title = {Information Revelation and Pandering in Elections},
  note = {unpublished},
  year = {2015},
  month = {April}
}

@article{CM85,
  title={Optimal Selling Strategies Under Uncertainty for a Discriminating Monopolist When Demands Are Interdependent},
  author={Cr{\'e}mer, Jacques and McLean, Richard P.},
  journal={Econometrica},
  volume={53},
  number={2},
  pages={345--361},
  year={1985},
}

@article{CM88,
  title={Full Extraction of the Surplus in Bayesian and Dominant Strategy Auctions},
  author={Cr{\'e}mer, Jacques and McLean, Richard P.},
  journal={Econometrica},
  volume={56},
  number={6},
  pages={1247--1257},
  year={1988},
}

@article{MR92,
  title={Correlated Information and Mechanism Design},
  author={McAfee, R. Preston and Reny, Philip J.},
  journal={Econometrica},
  volume={60},
  number={2},
  pages={395--421},
  year={1992},
}

@book{Durrett95,
	Author = {Richard Durrett},
	Edition = {2nd},
	Publisher = {Duxbury Press},
	Title = {Probability: Theory and Examples},
	Year = {1995}}

@article{KLR23,
    author = {Kartik, Navin and Lee, SangMok and Rappoport, Daniel},
    title = {Single-Crossing Differences in Convex Environments},
    journal = {Review of Economic Studies},
    volume = {91},
    number = {5},
    pages = {2981--3012},
    year = {2024},
    doi = {10.1093/restud/rdad103},
}

@article{QS12,
	author = {Quah, John K-H and Strulovici, Bruno},
	journal = {Econometrica},
	month = {September},
	number = {5},
	pages = {2333--2348},
	title = {Aggregating the Single Crossing Property},
	volume = {80},
	year = {2012}}

\onehalfspacing
\newpage %

\begin{center}
\textbf{{\LARGE {Supplementary Appendix}}}
\end{center}

\section{Mixed Motives}
\label{app:mixed}

This section substantiates the discussion in \autoref{sec:elections-discussion} of the paper
by formally generalizing our main welfare conclusions to a normal-quadratic
setting in which candidates are largely but not entirely office motivated.
We will establish that when the parameters $b_{i}$ and $\rho _{i}$ defined
in \autoref{e:mixed-motivations} are sufficiently close to zero for each $%
i=A,B$, 
(i) there is an equilibrium that achieves welfare arbitrarily close to the
level obtained by efficiently aggregating the signal of only one candidate (%
\autoref{p:UD-ideology} below), and (ii) that welfare is an approximate
bound on voter welfare in any equilibrium within a class (\autoref{thm:ideology-robustness}
below).

In the context of a normal-quadratic mixed-motivation game, with candidates'
payoffs as defined in \autoref{e:mixed-motivations}, we say that candidate 
$i$'s strategy is \emph{unbiased} if 
\begin{equation}
y_{i}(s_{i})=\frac{\beta }{\alpha +\beta }s_{i}+b_{i}.  \label{e:UD-ideology}
\end{equation}%
Note that this refers to candidate $i$ choosing a policy that maximizes 
\emph{his} preference over policy given his signal, as opposed to the
voter's.

\begin{proposition}
\label{p:UD-ideology} In the normal-quadratic mixed-motivations game, there
is a fully revealing equilibrium in which one candidate $i$ plays the
unbiased strategy \eqref{e:UD-ideology}, the other candidate $-i$ plays 
\begin{equation}
y_{-i}(s_{-i})=s_{-i}-\frac{\alpha +\beta }{\beta }b_{i},
\label{e:UD-ideology-loser}
\end{equation}%
and the voter elects candidate $i$ no matter the pair of platforms. 
\end{proposition}

\begin{proof}
Given the strategies \eqref{e:UD-ideology} and \eqref{e:UD-ideology-loser},
it follows that 
\begin{equation*}
\mathbb{E}[\theta  \mid x_{i},x_{-i}]=\frac{\beta (x_{i}-b_{i})\frac{\alpha
+\beta }{\beta }+\beta \left( x_{-i}+\frac{\alpha +\beta }{\beta }%
b_{i}\right) }{\alpha +2\beta }=\frac{\alpha x_{i}+\beta (x_{i}+x_{-i})}{%
\alpha +2\beta }.
\end{equation*}%
Straightforward algebra then verifies that
\[
\left( x_{-i}-\mathbb{E}[\theta \mid x_{i},x_{-i}]\right) ^{2}-\left( x_{i}-\mathbb{E}[\theta \mid x_{i},x_{-i}]\right) ^{2}=\frac{\alpha }{\alpha +2\beta }\left( x_{i}-x_{-i}\right) ^{2}\geq 0,
\]
with strict inequality whenever $x_{i}\neq x_{-i}$. Hence it is optimal for the voter to always elect candidate $i$; clearly the candidates are playing optimally given this strategy for the voter.
\end{proof}

As the equilibrium constructed in \autoref{p:UD-ideology} is invariant to $%
\rho_A$ and $\rho_B$, it has a number of interesting implications. First,
the equilibrium exists when candidates are purely policy-motivated. Second,
for $\rho_A=\rho_B=b_A=b_B=0$, this equilibrium reduces to one that verifies
the first statement of \autoref{p:UD}. 
Moreover, by taking $b_A=b_B=0$ and $\rho_A=\rho_B=1$, we see that there is
also an equilibrium in which one candidate plays the unbiased strategy and
always wins when both candidates are benevolent. 
Hence, the equilibrium of \autoref{p:UD-ideology} continuously spans all
three polar cases of candidate motivation.

Consider a normal-quadratic game with mixed-motivated candidates
parameterized by $(\bm{\rho},\bm{b})$, where $\bm{\rho}\equiv (\rho
_{A},\rho _{B})$ and $\bm{b}\equiv (b_{A},b_{B})$. Let $\mathcal{E}(%
\bm{\rho,b})$ denote the set of equilibria in which candidates play pure
strategies, for consistency with our baseline model. Given any equilibrium $%
\sigma\equiv (y_A,y_B,w_A)$, let $v(\sigma)$ be the voter's welfare in this
equilibrium. Note that the voter's welfare depends only on the strategies
used and not directly on the candidates' motivations. 
Let $v^*(\bm{\rho,b}):=\sup \{v(\sigma): \sigma \in \mathcal{E}(\bm{\rho,b}%
)\}$ be the supremum of equilibrium voter welfare given candidate
motivations. Plainly, $v^*(\bm{0,0})$ is the welfare bound identified by %
\autoref{thm:competition}.

For $C>0$, let $\mathcal{E}^{\uparrow}_{C}(\bm{\rho,b})\subseteq \mathcal{E}(\bm{\rho,b})$ denote
the set of equilibria in which each candidate $i$'s strategy $y_{i}$ is nondecreasing and
satisfies the linear growth bound
\begin{equation}
\label{e:growth}
\abs{y_{i}(s_{i})}\leq C\left( 1+\abs{s_{i}}\right) \quad \text{for all }s_{i},
\end{equation}
and let $v^{\uparrow}_{C}(\bm{\rho,b}):=\sup \{v(\sigma ):\sigma \in \mathcal{E}^{\uparrow}_{C}(%
\bm{\rho,b})\}$. The strategies in \autoref{p:UD-ideology} are affine and strictly increasing, with
coefficients bounded and intercepts vanishing as $(\bm{\rho,b})\rightarrow (\mathbf{0},\mathbf{0}%
)$; so there is $C_{0}>0$ such that $\mathcal{E}^{\uparrow}_{C}(\bm{\rho,b})$ is nonempty for every
$C\geq C_{0}$ and all $(\bm{\rho,b})$ in a neighborhood of $(\mathbf{0},\mathbf{0})$.

\begin{proposition}
\label{thm:ideology-robustness} Fix an arbitrary $C\geq C_{0}$. In the normal-quadratic
mixed-motivations game, $v^{\uparrow}_{C}(\bm{\rho,b})\rightarrow v^{*}(\mathbf{0},\mathbf{0})$ as
$(\bm{\rho},\bm{b}) \rightarrow (\mathbf{0},\mathbf{0})$.
\end{proposition}

The proposition restricts attention to equilibria in nondecreasing strategies satisfying \eqref{e:growth}. Monotonicity of strategies is  standard; the role of \eqref{e:growth} is to ensure the compactness and uniform integrability used in the proof. Any conditions delivering those properties would suffice; but we have not been able to dispense with them using welfare maximization considerations alone.\footnote{To see that welfare maximization does not imply the growth bound \eqref{e:growth}, consider a variant of the equilibrium of \autoref{p:UD-ideology} in which the losing candidate plays a constant strategy. Choosing a sequence of constants that diverge as $(\bm{\rho,b})\rightarrow (\mathbf{0},\mathbf{0})$ yields equilibria with the same welfare as those of \autoref{p:UD-ideology} yet violate \eqref{e:growth} for every fixed $C$.}

\begin{proof}[\textbf{Proof of \autoref{thm:ideology-robustness}}]
That $v^{*}(\mathbf{0},\mathbf{0})$ is a lower bound at the limit is immediate: the equilibrium $\sigma^{\bm{\rho,b}}_{\text{UB}}$ of
\autoref{p:UD-ideology} lies in $\mathcal{E}^{\uparrow}_{C}(\bm{\rho,b})$, so $v^{\uparrow}_{C}(%
\bm{\rho,b})\geq v(\sigma^{\bm{\rho,b}}_{\text{UB}}) \rightarrow v^{*}(\mathbf{0},\mathbf{0})$.

To show that $v^{*}(\mathbf{0},\mathbf{0})$ is also an upper bound at the limit, fix any sequence $(\bm{\rho}^{n},\bm{b}^{n})\rightarrow (\mathbf{0},\mathbf{0}%
)$. For each $n$ choose $\sigma ^{n}\equiv (y^{n}_{A},y^{n}_{B},w^{n}_{A})\in \mathcal{E}
^{\uparrow}_{C}(\bm{\rho}^{n},\bm{b}^{n})$ with $v(\sigma ^{n})\geq v^{\uparrow}_{C}(\bm{\rho}^{n},%
\bm{b}^{n})-1/n$, and pass to a subsequence along which $v(\sigma ^{n})$
converges to $\limsup_{n}v^{\uparrow}_{C}(\bm{\rho}^{n},\bm{b}^{n})$. Every further subsequence
extracted below preserves that limit, so it suffices to show $\lim_{n}v(\sigma ^{n})\leq
v^{*}(\mathbf{0},\mathbf{0})=-1/(\alpha +\beta )$.

\textbf{Step 1:} We first show that the candidates' strategies converge. By \eqref{e:growth}, the $y^{n}_{i}$ are uniformly bounded on each compact interval. Since they are also nondecreasing, Helly's selection theorem and a diagonal argument yield a subsequence along which $y^{n}_{i}$ converges at every continuity point of a nondecreasing limit $y_{i}$, hence $F_{s_i}$-a.e., as $F_{s_i}$ is atomless. Because $y_{i}$ inherits the bound \eqref{e:growth}, we have $\abs{y^{n}_{i}-y_{i}}^{2}\leq 4C^{2}(1+\abs{s_{i}})^{2}$, which is $F_{s_i}$-integrable, so dominated convergence gives $y^{n}_{i}\rightarrow y_{i}$ in $L^{2}(F_{s_i})$.

\textbf{Step 2:} Next, we establish a version of incentive compatibility at the limit, inequality \eqref{e:limit-ic} below. Define the probability that $A$ wins at a signal profile,
\begin{equation*}
q_{n}(s_{A},s_{B}):=w^{n}_{A}\!\left( y^{n}_{A}(s_{A}),y^{n}_{B}(s_{B})\right),
\end{equation*}
and let $\lambda :=F_{s_A}\otimes F_{s_B}$. Since $F_{s_A,s_B}$ and $\lambda$ have everywhere-positive densities on $\Reals^{2}$, they have the same null sets. As $0\leq q_{n}\leq 1$ and $L^{1}(\lambda )$ is separable, a further subsequence satisfies $q_{n}\overset{\ast }{\rightharpoonup }q$ in
$L^{\infty }(\lambda )$,\footnote{That is, we have weak$^*$ convergence: $\int q_{n}h\,\mathrm d\lambda \rightarrow \int qh\,\mathrm
d\lambda$ for every $h\in L^{1}(\lambda )$.} with $0\leq q\leq 1$ $\lambda$-a.e.

Let us now pass the candidates' incentive constraints to the limit. In $\sigma ^{n}$, candidate $i$ with signal $s_{i}$ may misreport any signal $t_i$, i.e., deviate to the platform $y^{n}_{i}(t_{i})$. Write $q^{A}_{n}(s_{A},s_{B}):=q_{n}(s_{A},s_{B})$ and $q^{B}_{n}(s_{B},s_{A}):=1-q_{n}(s_{A},s_{B})$, and let $H_{n,i}(s_{i},t_{i}):=\E[q^{i}_{n}(t_{i},s_{-i})\mid s_{i}]$. For all sufficiently large $n$ we have $\rho ^{n}_{i}<1$.  Dividing candidate $i$'s equilibrium incentive-compatibility condition by $1-\rho ^{n}_{i}$, we have that for $F_{s_i}$-a.e.~$s_{i}$ and every $t_{i}$,
\begin{equation*}
H_{n,i}(s_{i},s_{i})\geq H_{n,i}(s_{i},t_{i})-\varepsilon ^{n}_{i}\,K\!\left(1+s_{i}^{2}+t_{i}^{2}\right) ,\qquad \varepsilon ^{n}_{i}:=\frac{\rho ^{n}_{i}}{1-\rho ^{n}_{i}}
\rightarrow 0,
\end{equation*}
where $K(1+s_{i}^{2}+t_{i}^{2})$ bounds the absolute difference between the conditional expected policy losses under truthful play and under the deviation, with $K$ uniform in $n$.\footnote{Writing $\Pi_{n,i}(s_{i},t_{i})$ for type $s_{i}$'s conditional expected policy loss from reporting $t_{i}$, and $\eta^{n}_{i}:=\theta +b^{n}_{i}$, the fact that election probabilities lie in $[0,1]$ implies
\[
\abs{\Pi_{n,i}(s_{i},s_{i})-\Pi_{n,i}(s_{i},t_{i})}\leq \E\!\left[ (y^{n}_{i}(s_{i})-\eta^{n}_{i})^{2}+(y^{n}_{i}(t_{i})-\eta^{n}_{i})^{2}+2(y^{n}_{-i}(s_{-i})-\eta^{n}_{i})^{2}\,\Big|\,s_{i}\right] \leq K(1+s_{i}^{2}+t_{i}^{2}),
\]
where the second inequality uses \eqref{e:growth}, the boundedness of $\bm{b}^{n}$, and the linear conditional means and bounded conditional variances of the normal specification.} Multiply the inequality by an arbitrary bounded nonnegative measurable $\psi (s_{i},t_{i})$ and integrate with respect to $F_{s_i}\otimes F_{s_i}$. As $n\rightarrow \infty$, the two election-probability terms converge, each being an integral of $q_{n}$ against a density in $L^{1}(\lambda )$, while the error term vanishes because $\varepsilon^{n}_{i}\rightarrow 0$ and $1+s_{i}^{2}+t_{i}^{2}$ is integrable. Since $\psi$ was arbitrary, we obtain, with $H_{i}$ defined from $q$ as above,
\begin{equation}
\label{e:limit-ic}
H_{i}(s_{i},s_{i})\geq H_{i}(s_{i},t_{i})\qquad \text{for }F_{s_i}\otimes F_{s_i}\text{-a.e.~}%
(s_{i},t_{i}).
\end{equation}

\textbf{Step 3:}
We claim \eqref{e:limit-ic} forces the limit winning probability function $q$ to be constant, even though \eqref{e:limit-ic} holds only for almost every report pair. Let $\bar q:=\int q\,\mathrm dF_{s_A,s_B}$ and $q^{\times }:=\int q\,\mathrm d\lambda$. Integrating \eqref{e:limit-ic} for candidate $A$ over $(s_{A},t_{A})$ gives $\bar q\geq q^{\times }$, and for candidate $B$ gives $1-\bar q\geq 1-q^{\times }$. So $\bar q=q^{\times }$, and the nonnegative integrand of \eqref{e:limit-ic} has zero integral and hence vanishes, so that $H_{i}(s_{i},s_{i})=H_{i}(s_{i},t_{i})$ a.e. In particular, for $F_{s_A}\otimes F_{s_A}$-a.e.~$(t, t^{\prime })$ we have $\E[\,q(t,s_{B})-q(t^{\prime },s_{B})\mid s_{A}\,]=0$ for $F_{s_A}$-a.e.~$s_{A}$. Applying \autoref{completeness_general} to $s_{B}\mapsto q(t,s_{B})-q(t^{\prime },s_{B})$ gives $q(t,\cdot )=q(t^{\prime },\cdot )$ $F_{s_B}$-a.s., i.e., $q$ is a.e.~independent of its first argument. The symmetric argument for candidate $B$ makes it a.e.~independent of its second. Hence $q$ equals some constant $c\in [0,1]$ $\lambda$-a.e., and therefore also $F_{s_A,s_B}$-a.e. Note that this never requires $q$ to be a best response to the limiting strategies $(y_{A},y_{B})$; the candidates' incentives alone force constancy.

\textbf{Step 4:} Finally, we pass welfare to the limit. Let $L^{n}_{i}(s_{A},s_{B}):=\E[(y^{n}_{i}(s_{i})-\theta )^{2}\mid s_{A},s_{B}]$, so that $-v(\sigma ^{n})=\E[L^{n}_{B}+q_{n}(L^{n}_{A}-L^{n}_{B})]$, and define $L_{i}$ analogously from $y_{i}$. Since the posterior mean of $\theta$ is linear in $(s_{A},s_{B})$ with finite second moment, $y^{n}_{i}\rightarrow y_{i}$ in $L^{2}(F_{s_i})$ implies $L^{n}_{i}\rightarrow L_{i}$ in $L^{1}(F_{s_A,s_B})$. As $0\leq q_{n}\leq 1$, it follows that $\E[L^{n}_{B}+q_{n}(L^{n}_{A}-L^{n}_{B})]$ and $\E[L_{B}+q_{n}(L_{A}-L_{B})]$ have the same limit. The latter converges by weak$^*$ convergence, since $r(L_{A}-L_{B})\in L^{1}(\lambda )$ for $r:=\mathrm dF_{s_A,s_B}/\mathrm d\lambda$. Hence, using $q=c$ (from the previous step),
\begin{equation*}
-\lim_{n} v(\sigma ^{n})=c\,\E\!\left[ (y_{A}(s_{A})-\theta )^{2}\right] +(1-c)\,\E\!\left[
(y_{B}(s_{B})-\theta )^{2}\right] .
\end{equation*}%
Each $y_{i}$ depends only on $s_{i}$, so each expectation is at least $\E[\Var(\theta \mid s_{i})]=1/(\alpha +\beta )$, and $\lim_{n}v(\sigma ^{n})\leq -1/(\alpha +\beta )=v^{*}(\mathbf{0},\mathbf{0})$.
\end{proof}

\section{A Beta-Bernoulli Specification}
\label{app:beta-bernoulli}

Here we repeat the analysis of \autoref{sec:pandering} for the case in which the state follows a Beta distribution and each candidate gets a binary signal drawn from a Bernoulli distribution. This statistical structure is a member of the exponential family with conjugate priors. Aside from illustrating how the incentives to overreact exist even when the state distribution may not be unimodal and may be skewed, signals are discrete, etc., it also provides a closer comparison with the setting of \citet{HL:03} and \citet{Loertscher:10} than does our leading normal-normal specification.

Assume the prior distribution of $\theta $ is $\Be(\alpha ,\beta )$, which is the Beta distribution with parameters $\alpha ,\beta >0$ whose density is given by $f\left( \theta \right) =\frac{\theta ^{\alpha -1}\left( 1-\theta\right) ^{\beta -1}}{B\left( \alpha ,\beta \right) }$, where $B(\cdot ,\cdot)$ is the Beta function.\footnote{If $\alpha $ and $\beta $ are positive integers then $B\left( \alpha ,\beta\right) =\frac{\left( \alpha -1\right) !\left( \beta -1\right) !}{\left(\alpha +\beta -1\right) !}$.} Thus $\theta $ has support $\left[ 0,1\right] $ and $\mathbb{E}[\theta ]=\frac{\alpha }{\alpha +\beta }$. For reasons explained at the end of the section, we assume $\alpha \neq \beta $. (This rules out a uniform prior, which corresponds to $\alpha =\beta =1$.) Each candidate $i\in \left\{ A,B\right\} $ observes a private signal $s_{i}\in \{0,1\}$; conditional on $\theta $, signals are drawn independently from the same Bernoulli distribution with $\Pr (s_{i}=1 \mid \theta )=\theta $. The policy space is any subset of $\mathbb{R}$ containing $[0,1]$.

It is well-known that the posterior distribution of the state given signal $1 $ is now $\Be\left( \alpha +1,\beta \right) $ (i.e., has density $f\left(\theta  \mid s_{i}=1\right) =\frac{\theta ^{\alpha }\left( 1-\theta \right)^{\beta -1}}{B\left( \alpha +1,\beta \right) }$); similarly the posterior given signal $0$ is $\Be\left( \alpha ,\beta +1\right) $. It is also straightforward to check that the posterior distribution of the state given two signals is as follows: if both $s_{i}=s_{-i}=1$, it is $\Be\left( \alpha+2,\beta \right) $; if $s_{i}=0$ and $s_{-i}=1$, it is $\Be\left( \alpha+1,\beta +1\right) $; and if $s_{i}=s_{-i}=0$, it is $\Be\left( \alpha,\beta +2\right) .$

It follows that 
\begin{equation*}
\mathbb{E}\left[ \theta  \mid s_{i}\right] =\frac{\alpha +s_{i}}{\alpha +\beta +1}%
\text{ and }\mathbb{E}\left[ \theta  \mid s_{i},s_{-i}\right] =\frac{\alpha
+s_{i}+s_{-i}}{\alpha +\beta +2}.
\end{equation*}

The above formulae imply that for any realization $(s_{A},s_{B})$, 
\begin{align}
\sign\left( \mathbb{E}[ \theta  \mid s_{A},s_{B}] -\mathbb{E}\left[ \theta \right]\right) & =\sign\left( \frac{\mathbb{E}\left[ \theta  \mid s_{A}\right] +\mathbb{E}\left[ \theta  \mid s_{B}\right] }{2}-\mathbb{E}\left[ \theta \right] \right) , 
\notag \\
\left\vert \mathbb{E}[ \theta  \mid s_{A},s_{B}] -\mathbb{E}\left[ \theta \right]\right\vert & >\left\vert \frac{\mathbb{E}\left[ \theta  \mid s_{A}\right] +\mathbb{E}\left[ \theta  \mid s_{B}\right] }{2}-\mathbb{E}\left[ \theta \right]\right\vert .  \label{e:BB}
\end{align}
Hence, both the posterior mean given two signals and the average of
the individual posterior means shift in the same direction from the prior
mean, but the former does so by more.

Consequently, if candidates were to play unbiased strategies and the voter
best responds, then whenever $s_{A}\neq s_{B}$ there is one candidate who
wins with probability one: the candidate $i$ with $s_{i}=1$ (resp., $s_{i}=0$) when $\beta >\alpha $ (resp., $\beta <\alpha $). Of course, when $s_{A}=s_{B}$, both candidates would choose the same platform and win with
equal probability. It is worth highlighting that when $s_{A}\neq s_{B}$, it
is the candidate with the ex-ante \emph{less} likely signal who wins,
because ex-ante $\Pr (s_{i}=1)=\mathbb{E}[\theta ]=\alpha /(\alpha +\beta )$. This implies that unbiased strategies cannot form an equilibrium, but not
because candidates would deviate when drawing the ex-ante less likely
signal; rather, they would deviate when drawing the ex-ante \emph{more}
likely signal to the platform corresponding to the ex-ante less likely
signal.\footnote{%
See \citet{CDK:11} for an analog where options that are \textquotedblleft
unconditionally better-looking\textquotedblright\ need not be
\textquotedblleft conditionally better-looking\textquotedblright\ .} Notice
that this profitable deviation given signal $s_{i}$ is to an (on-path)
platform $x_{i}$ such that $\abs{x_{i}-\mathbb{E}[\theta ]}>\abs{\mathbb{E}[\theta 
\mid s_{i}]-\mathbb{E}[\theta ]}$; hence, it is a profitable deviation through
overreaction rather than pandering.

Finally, we observe that there is a symmetric fully revealing equilibrium with
overreaction in which both candidates play 
\begin{equation*}
y(1)=\frac{\alpha +2}{\alpha +\beta +2} \quad \text{and} \quad y(0)=\frac{\alpha }{\alpha +\beta +2}.
\end{equation*}
This strategy displays overreaction because 
\begin{equation*}
y(0)<\mathbb{E}[\theta  \mid s_{i}=0]<\mathbb{E}[\theta ]<\mathbb{E}[\theta
 \mid s_{i}=1]<y(1).
\end{equation*}
It is readily verified that when both candidates use this strategy, $\mathbb{E}[ \theta  \mid s_{A},s_{B}] =\frac{y(s_{A})+y(s_{B})}{2}$ for all $(s_{A},s_{B}) $, and hence each candidate would win with probability $1/2$
for all on-path platform pairs; a variety of off-path beliefs can be used to
support the equilibrium.

Note that this overreaction equilibrium would exist even when $\alpha =\beta 
$. However, were $\alpha =\beta $, unbiased strategies would also constitute
an equilibrium: for, given unbiased strategies, both sides of \eqref{e:BB}
would be equal to each other (in fact, equal to zero) when $s_{A}\neq s_{B}$, and hence the voter could elect both candidates with equal probability for all on-path platform pairs; again, suitable off-path beliefs support the equilibrium.
\end{document}